\documentclass{elsart}
\usepackage{graphicx}
\usepackage{xspace}
\usepackage{amssymb}
\usepackage{amsmath}
\usepackage{latexsym}
\usepackage{mathrsfs}

\newcommand{\diff}{{\rm\,d}}
\newcommand{\bint}{\mskip .5mu \int \mskip-18mu}

\def\s{\mbox{\boldmath $s$}}

\def\p{\mbox{\boldmath $p$}}

\def\q{\mbox{\boldmath $q$}}

\def\k{\mbox{\boldmath $k$}}

\def\r{\mbox{\boldmath $r$}}
\def\rf{\mbox{\boldmath $r^{ \prime}$}}
\def\pf{\mbox{\boldmath $p^{ \prime}$}}
\def\sf{\mbox{\boldmath $s^{ \prime}$}}

\def\mcg{\mbox{$\mathcal{G}$}}
\def\mcv{\mbox{$\mathcal{V}$}}
\def\mcs{\mbox{$\mathcal{S}$}}
\def\mct{\mbox{$\mathcal{T}$}}
\def\mcm{\mbox{$\mathcal{M}$}}
\def\mch{\mbox{$\mathcal{H}$}}
\def\mcu{\mbox{$\mathcal{U}$}}
\def\mcd{\mbox{$\mathcal{D}$}}
\begin{document}
\begin{frontmatter}
\title{Antisymmetrized Green's function approach to $(e,e')$ reactions
with a realistic nuclear density}

\author[pv]{F.~Capuzzi},
\author[pv]{C.~Giusti},
\author[pv]{F.~D.~Pacati}
\address[pv]{Dipartimento di Fisica Nucleare e Teorica,
Universit\`{a} di Pavia and \\
Istituto Nazionale di Fisica Nucleare, Sezione di Pavia, I-27100
Pavia, Italy}
\author[so]{D.~N.~Kadrev}
\address[so]{Institute for Nuclear Research and Nuclear Energy,
Sofia 1784, Bulgaria}

\date{\today}

\begin{abstract}
A completely antisymmetrized Green's function approach to the
inclusive quasielastic $(e,e')$ scattering, including a realistic
one-body density, is presented. The single particle Green's function
is expanded in terms of the eigenfunctions of the nonhermitian
optical potential. This allows one to treat final state interactions
consistently in the inclusive and in the exclusive reactions.
Nuclear correlations are included in the one-body density. Numerical
results for the response functions of $^{16}$O and
$^{40}$Ca are presented and discussed.
\end{abstract}


\end{frontmatter}

\section{Introduction \label{sec.intro}}

The one-body mechanism gives a natural interpretation of the
inclusive electron scattering in the quasielastic region. However,
in order to explain the experimental data of the separated
longitudinal and transverse responses more complicated mechanisms
are needed. A review of the experimental data and their possible
explanations can be found in Ref. \cite{book}. Thereafter, only a few
experimental papers were published \cite{batesca,csfrascati}, while
new experiments with high resolution are planned at JLab
\cite{jlabpro}.

Many papers were published in order to explain the problems raised by
the separation, i.e., the apparent lack of strength in the
longitudinal response and the apparent excess of strength in the
transverse one. Among them, the more recent ones are concerned with
the contribution to the inclusive cross section of meson exchange
currents and isobar excitations \cite{Sluys,Cenni,Amaro,Amaro2}, with
the effect of correlations \cite{Fabrocini,Co}, and the use of a
relativistic framework in the calculations \cite{Amaro,meucci}.

At present, however, a consistent and simultaneous description of the
longitudinal and transverse response functions is not available. A
possible solution could be the combined effect of two-body currents
and tensor correlations \cite{Leidemann,Fabrocini,Sick}.

A peculiar problem of inclusive electron scattering is the treatment
of final state interactions (FSI), since they are essential to
explain the exclusive reaction with one-nucleon emission, which is
the dominant process contributing to the inclusive reaction in the
quasielastic region. The large absorption due, e.g., to the
imaginary part of the optical potential, produces a loss of flux
that is appropriate for the exclusive process, but inconsistent for
the inclusive one, where the total flux must be conserved. This
conservation is preserved in the Green's function approach considered
here. This result was originally derived by arguments based on the
multiple scattering theory \cite{hori} and then by means of the
Feshbach projection operator formalism
\cite{chinn,bouch,capuzzi,capma,meucci}.

The spectral representation of the single-particle (s.p.) Green's
function, based on a biorthogonal expansion in terms of the
eigenfunctions of the nonhermitian optical potential, allows one to
perform explicit calculations and to treat FSI consistently in the
inclusive and in the exclusive reactions. In previous papers
\cite{capuzzi,meucci} the Green's function approach was used both in
a nonrelativistic and in a relativistic framework to perform explicit
calculations of the inclusive response functions.

Two issues are the main goal of this paper: a completely
antisymmetrized presentation of the Green's function approach, and
the effect of nuclear correlations, which are included in the model
through the use of a realistic one-body density matrix (ODM). In
the previous application of the method \cite{capuzzi,meucci}
correlations were neglected. The response functions were given by a
sum over the residual nucleus states restricted to be s.p. one-hole
states in the target. A pure shell model was assumed for the nuclear
structure and therefore the overlap functions between the target and
the residual nucleus were given by phenomenological s.p. wave
functions with a shell-model spectroscopic factor. In the present
work we are able to include partial occupation numbers, and
therefore correlations, through the natural expansion of the ODM.

The definitions and the main properties of the quantities involved in
the model are given in Section~\ref{sec.def}. In Sec.~\ref{sec.green}
the antisymmetrized Green's function approach is developed and the
inclusive cross section is expressed in terms of the ODM. In
Sec.~\ref{sec.spectral} the Green's function is calculated in terms of
the spectral representation related to the optical potential. 
In Sec.~\ref{sec.density} some models describing the ODM including 
correlations are briefly reviewed.
In Sec.~\ref{sec.results} the results of calculations for $^{16}$O and
$^{40}$Ca are reported and compared with some experimental data.
Summary and conclusions are drawn in Sec.~\ref{conc}.

\section{Definitions and main properties \label{sec.def}}

In this Section we collect most of the definitions which will be
useful in the treatment of the inclusive $(e,e')$ reaction, using the
Green's functions approach along the lines of
Refs.~\cite{capuzzi,capma,meucci}.

\subsection{One-body density, Green's functions and related
quantities \label{ssec:2.1}}

We deal in this paper with three different systems: the $Z$--proton
residual nucleus, the ($Z+1$)--proton target nucleus, and the
($Z-1$)--proton system obtained removing a proton from the residual
nucleus. 
The contribution of the neutrons, here disregarded in order to simplify 
the formalism, can be introduced in an obvious way.
Let $T$ and $H$ be the kinetic energy and the Hamiltonian
operators acting in the Hilbert spaces
$\mathcal{H}^{Z},\mathcal{H}^{Z+1}$, and $\mathcal{H}^{Z-1}$,
corresponding to the three given systems and written in the Fock's
formalism. Here we are interested only in the following eigenvectors
of $H$: i) the eigenvectors $|n\rangle$, representing the bound
states of the residual nucleus with energy $\epsilon_n$, ii) the
eigenvector $|\psi_0\rangle$, representing the initial state, which
is the ground state of the target nucleus with energy $E_0$, and
iii) the eigenvectors $|\psi_f\rangle$, representing all the final
(bound or unbound) states of the target nucleus, including
$|\psi_0\rangle$. These eigenvectors are properly orthonormalized.
Relatedly, we define the hole overlaps, referred to $|\psi_0\rangle$,
\begin{equation} \label{eq:2.1}
\langle\r |\phi_n \rangle \equiv \langle n |a_{\mathbf {r}} |\psi_0
\rangle,
\end{equation}
and the particle overlaps, referred to $|n\rangle$,
\begin{equation}
\langle\r |\chi_{f,n} \rangle \equiv \langle n |a_{\mathbf {r}}
|\psi_f \rangle, \label{eq:2.2}
\end{equation}
where $a_{\mathbf {r}}$ annihilates a proton at the point $\r$. The
corresponding annihilation operator of a proton of momentum $\p$ is
\begin{equation}
a_{\mathbf{p}} \equiv (2\pi)^{-3/2} \int\diff\r \exp(-i\p \cdot \r) \
a_{\mathbf{r}}. \label{eq:2.3}
\end{equation}
For sake of simplicity, we consider spinless protons, we assume that
$|\psi_0 \rangle$ and $|n\rangle$ are not degenerate, and we omit any
degeneracy index in the unbound states $|\psi_f \rangle$. When
referring to quantities related to $n = 0$, the index 0 will be
usually suppressed in the following Sections.

\subsubsection{The one-body density matrix}

The one-body density matrix related to $|\psi_0 \rangle$ is defined
as
\begin{equation}
\langle\r|K|\rf\rangle = \langle\psi_0| a_{\mathbf {r'}}^{\dagger}
a_{\mathbf {r}}|\psi_0 \rangle . \label{eq:2.4}
\end{equation}
It is real and symmetric by the exchange $\r \longleftrightarrow
\rf$, since $|\psi_0 \rangle$ is not degenerate. The corresponding
operator $K$ in $L^2({\bf R}^3)$ is nonnegative, self-adjoint, and
has a finite trace equal to $Z$+1. The eigenvalue equation
\begin{equation}
K|u_\nu\rangle = n_\nu|u_\nu\rangle \label{eq:2.5}
\end{equation}
defines the natural orbitals $u_\nu(\r) = \langle\r|u_\nu\rangle$ and
the related occupation numbers $n_\nu$, which satisfy the relations
\begin{equation}
0 \leq n_\nu \leq 1 \, ,\, \, \sum_\nu n_\nu = Z + 1. \label{eq:2.6}
\end{equation}
The operator $K$ is invertible in the subspace of $L^2({\bf R}^3)$
spanned by the orbitals $u_\nu$, with $n_\nu \neq 0$. If $n_\nu = 0$
is not an eigenvalue of $K$, it is necessarily an accumulation point
of eigenvalues. Thus $K$ is fully invertible, but $K^{-1}$ is a
rather complicate operator.

\subsubsection{The one-body Green's function}

At complex energies $z$, the one-body Green's function related to
$|n\rangle$ is the sum of the particle and hole Green's functions:
\begin{equation}
\langle \r |\mcg_n(z)|\rf\rangle = \langle \r
|\mcg_n^{(p)}(z)|\rf\rangle + \langle \r |\mcg_n^{(h)}(z)|\rf\rangle
\label{eq:2.7}
\end{equation}
with
\begin{equation}
\langle \r |\mcg_n^{(p)}(z)|\rf\rangle = \langle n |a_{\r} (z - H +
\epsilon_n)^{-1} a_{\rf}^{\dagger} |n\rangle , \label{eq:2.8}
\end{equation}
\begin{equation}
\langle \r |\mcg_n^{(h)}(z)|\rf\rangle = \langle n |a_{\rf}^{\dagger}
(z + H - \epsilon_n)^{-1} a_{\r} |n\rangle . \label{eq:2.9}
\end{equation}
These Green's functions are symmetric by the exchange $\r
\longleftrightarrow \rf$, since $|n\rangle$ is not degenerate.

At real energies $E$, we consider the retarded Green's function
instead of the time-ordered one, since it is more convenient in the
formalism developed below. It is defined as
\begin{equation}
\mcg_n(E) = \lim_{\eta \rightarrow +0} \mcg_n(E+i\eta).
\label{eq:2.10}
\end{equation}
In this equation, as well as in all the next equations involving
$\eta$, the limit is understood in the weak sense, i.e., it must be
performed after inclusion into a scalar product between normalizable
vectors. Henceforth, the symbol of limit will be usually understood.

\subsubsection{The spectral function}

The Green's function $\mcg_n(z)$ is analytic in the complex plane
except for cuts and poles on the real axis. Hence, it can be written
as
\begin{equation}
\mcg_n(z) = \int_{-\infty}^{+\infty} \diff E' \, \frac {\mcs_n(E')}
{z-E'}, \label{eq:2.11}
\end{equation}
where
\begin{equation}
\mcs_n(E) = \frac {1} {2\pi i} [\mcg_n(E-i\eta) - \mcg_n(E+i\eta)] =
\frac {1} {2\pi i} [\mcg_n^{\dagger}(E) - \mcg_n(E)] \label{eq:2.12}
\end{equation}
is the spectral function, which satisfies the sum rule
\begin{equation}
\int_{-\infty}^{+\infty} \diff E \, \mcs_n(E) = 1. \label{eq:2.13}
\end{equation}
Since $\mcs_n(E)$ is nonnegative, Eq.~(\ref{eq:2.11}) and
(\ref{eq:2.13}) show that
\begin{equation}
\mcg_n(z)|\phi \rangle = 0 \, \, \Longrightarrow \, |\phi\rangle = 0
\end{equation}
for $z$ complex. Hence, $\mcg_n(z)$ is fully invertible.

\subsubsection{The self-energy}

At complex energies $z$, the Green's function can be written as
\begin{equation}
\mcg_n(z) = \frac {1} {z - \mct -\mcm_n(z)} , \label{eq:2.15}
\end{equation}
where $\mct$ is the one-body kinetic energy and $\mcm_n$ the one-body
self-energy. The related Hamiltonian $h_n(z)$ is defined as
\begin{equation}
h_n(z) \equiv \mct + \mcm_n(z) = z - [\mcg_n(z)]^{-1}.
\label{eq:2.14}
\end{equation}
Note that $\mcg_n(z)$ is fully invertible, so that $\mcm_n(z)$ has no
mathematical drawbacks related to undue restrictions of its domain.
In contrast, $\mcg_n^{(p)}(z)$ is only partially invertible if 1 is
an eigenvalue of the density matrix associated with $|n\rangle$. This
produces a restriction of the domain of the related self-energy
$\mcm_n^{(p)}(z)$ which may lead, e.g., to an incorrect Dyson
equation. The same drawback affects $\mcm_n^{(h)}(z)$ if 0 is an
eigenvalue of the density matrix.

The analyticity of $\mcg_n(z)$ induces similar analyticity properties
into $h_n(z)$ and $\mcm_n(z)$. They are analytic in the complex plane
except for cuts and poles on the real axis, which are different from
those of $\mcg_n(z)$.

The one-body self-energy at real energies $\mcm_n(E)$ is defined as
\begin{equation}
\mcm_n(E) = \lim_{\eta \rightarrow +0} \mcm_n(E+i\eta)
\label{eq:2.16}
\end{equation}
In the Appendix B of \cite{capma3} it is proved that $\mcg_n(E)$ is
fully invertible, except for the values of $E$ corresponding to the
poles of $\mcm_n(E)$, and that holds the relation
\begin{equation}
\mcg_n(E) = \frac {1} {E - \mct - \mcm_n(E) + i \eta} .
\label{eq:2.17}
\end{equation}
The related Hamiltonian $h_n(E)$ is defined as
\begin{equation}
h_n(E) \equiv \mct + \mcm_n(E) = E - [\mcg_n(E)]^{-1} .
\label{eq:2.17a}
\end{equation}

\subsection{Self-energy and extended projection operators
\label{ssec:2.2}}

In his pioneering papers~\cite{fesh,fesh2} Feshbach introduced a
projection operator formalism to obtain a closed expression for the
theoretical optical-model potential. Here, we use an analogous
procedure for the self-energy, based on different projection
operators. More details can be found in Ref.~\cite{capma3}.

Let us introduce the vectors $\alpha_{\mathbf {r}}|n \rangle$, where
\begin{equation}
\alpha_{\mathbf {r}} \equiv a_{\mathbf {r}} + a_{\mathbf
{r}}^{\dagger} \, . \label{eq:2.18}
\end{equation}
They belong to the Hilbert space $\mch^{(Z+1)} \oplus \mch^{(Z-1)}$
and form an orthonormal set. In fact one has
\begin{equation}
\langle n|\alpha_{\mathbf {r}} \alpha_{\mathbf {r'}}|n \rangle =
\langle n|a_{\mathbf {r}} a_{\mathbf {r'}}^{\dagger} + a_{\mathbf
{r'}}^{\dagger} a_{\mathbf {r}} |n \rangle = \delta (\r - \rf),
\label{eq:2.19}
\end{equation}
where we have used the property that $\langle n| a_{\mathbf{r}
}^{\dagger} a_{\mathbf {r'}}|n \rangle$ is symmetric since $|n
\rangle$ is not degenerate. Therefore the operators
\begin{equation}
P_n = \int \diff \r \, \alpha_{\mathbf {r}}|n \rangle \langle n|
\alpha_{\mathbf {r}} = \int \diff \p \, \alpha_{\mathbf {p}} |n
\rangle \langle n| \alpha_{\mathbf {p}} \ , \, \, Q_n = 1 - P_n
\label{eq:2.20}
\end{equation}
are projection operators in $\mch^{(Z+1)} \oplus \mch^{(Z-1)}$. Such
operators, which will have an important role in Sec.~\ref{sec.green},
are called {\lq\lq extended projection operators\rq\rq} in order to
distinguish them from the Feshbach's ones. For sake of simplicity, we
define here $P_n$ in the simpler case of spinless particles. The
necessary changes to include spin and isospin variables can be found
in the Appendix A of Ref.~\cite{capma3}.

One can easily check that the correspondence
\begin{equation}
P_n |\phi \rangle \longleftrightarrow \langle n| \alpha_{\mathbf {r}}
|\phi \rangle , \label{eq:2.22}
\end{equation}
\begin{equation}
P_n O P_n \longleftrightarrow \langle n| \alpha_{\mathbf {r}} O
\alpha_{\mathbf {r'}}|n \rangle , \label{eq:2.23}
\end{equation}
where $|\phi \rangle$ and $O$ are vectors and operators in
$\mch^{(Z+1)} \oplus \mch^{(Z-1)}$, defines an isomorphism between
the Hilbert space of the vectors $P_n |\phi \rangle$
and $L^2({\bf R}^3)$. This means that every property
involving vectors and operators of one space implies that the same
property holds in the other space.

The above defined isomorphism is useful to develop a more synthetic
formalism for the quantities defined in Sec.~\ref{ssec:2.1}. To this
extent, we introduce the Hamiltonian-type operator $\hat H_n$, defined
by the relations
\begin{equation}
\hat H_n = H - \epsilon_n \quad \mathrm{in} \quad \mch^{(Z+1)} \quad
, \quad \hat H_n = \epsilon_n - H \quad \mathrm{in} \quad
\mch^{(Z-1)} , \label{eq:2.24}
\end{equation}
and extended to the whole space $\mch^{(Z+1)} \oplus \mch^{(Z-1)}$ by
linearity. Moreover, we introduce the related quantities
\begin{eqnarray}
\hat G_n(z) & = & \frac {1} {z - \hat H_n} \quad , \quad \hat G_n(E)
=
\hat G_n(E+i\eta) , \nonumber \\
\hat S_n(E) & = & \delta(E - \hat H_n) = \frac {1} {2\pi i} (\hat
G_n^{\dagger}(E) - \hat G_n(E)). \label{eq:2.25}
\end{eqnarray}
By the definition of $\delta(E - \hat H_n)$ one has
\begin{equation}
\int _{-\infty}^{+\infty} \diff E \, \hat S_n(E) = 1. \label{eq:2.26}
\end{equation}

Due to the orthogonality between vectors related to different numbers
of protons and due to the symmetry properties of the particle and
hole components of $\mcg_n(z)$, Eq.~(\ref{eq:2.7}) is reduced to the
simple expression
\begin{equation}
\langle \r |\mcg_n(z)|\rf\rangle = \langle n |\alpha_{\mathbf {r}}
\hat G_n(z) \alpha_{\mathbf {r'}}|n\rangle . \label{eq:2.27}
\end{equation}
Thus, $\mcg_n(z)$ is the one-body operator isomorphically
corresponding to the many-body operator $P_n\hat G_n(z)P_n$. 
Due to Eqs.~(\ref{eq:2.12}) and (\ref{eq:2.25}) it follows that $\mcs_n(E)$
corresponds to $P_n\delta(E-\hat H_n)P_n$. Therefore, one has
\begin{equation}
\langle \r |\mcs_n(E)|\rf\rangle = \langle n |\alpha_{\mathbf {r}}
\delta (E - \hat H_n) \alpha_{\mathbf {r'}}|n\rangle . \label{eq:2.28a}
\end{equation}
As the
handling of equations is often simpler when one deals with many-body
operators of this type, we introduce, in keeping with the definitions
of Sec.~\ref{ssec:2.1}, the many-body definitions of the Green's
function, the spectral function and the self-energy:
\begin{eqnarray}
G_n(z) & =& P_n\hat G_n(z)P_n \quad , \quad G_n(E) = P_n\hat
G_n(E)P_n \, , \\
S_n(E) & =& P_n\hat S_n(E)P_n = P_n \, \delta(E - \hat H_n)P_n =
\frac{1}{2\pi i} [G_n^{\dagger}(E) - G_n(E)], \\
h_n(z) & = & P_n\left(z - [\hat G_n(z)]^{-1}\right) P_n \, , \,
h_n(E) = P_n\left(E - [\hat G_n(E)]^{-1}\right) P_n \, , \\
M_n(z) & = & h_n(z) - P_n \hat T_n P_n \quad , \quad M_n(E) =
h_n(E) - P_n \hat T_n P_n \, , \label{eq:2.31}
\end{eqnarray}
where
\begin{equation}
\hat T_n \equiv \int \diff \k \, \alpha_{\mathbf {k}} |n \rangle
\frac {\k^2} {2m} \langle n |\alpha_{\mathbf {k}} , \label{eq:2.32}
\end{equation}
is the many-body operator which corresponds isomorphically to the
one-body kinetic energy $\mct$.

The closed expression of the many-body self-energy is deduced in
Secs.~3.1--3.3 of Ref.~\cite{capma3} and is:
\begin{equation}
M_n(z) = P_n[\hat H_n - \hat T_n + \hat H_n Q_n (z - Q_n \hat H_n
Q_n)^{-1} Q_n \hat H_n]P_n \, . \label{eq:2.33}
\end{equation}
For a two-body potential
\begin{equation}
V = \frac {1} {4} \int \diff\r \diff\s \diff\rf \diff\sf \,
a_{\mathbf {r}}^{\dagger} a_{\mathbf {s}}^{\dagger} {\mathcal V}
(\r,\s,\rf,\sf) a_{\mathbf {s'}} a_{\mathbf {r'}} , \label{eq:2.34}
\end{equation}
the corresponding one-body self-energy $\mcm_n(z)$ is the sum of the
static and dynamical parts
\begin{equation}
\langle \r| \mcm_n^{(S)}|\rf \rangle =\langle n|J_{\mathbf {r}}
a_{\mathbf {r'}}^{\dagger} + a_{\mathbf {r'}}^{\dagger} J_{\mathbf
{r}}|n\rangle , \label{eq:2.35}
\end{equation}
\begin{equation}
\langle \r| \mcm_n^{(D)}|\rf \rangle =\langle n|(J_{\mathbf {r}} +
J_{\mathbf {r}}^{\dagger})Q_n (z - Q_n \hat H_n Q_n)^{-1}
Q_n(J_{\mathbf {r'}} + J_{\mathbf {r'}}^{\dagger})|n\rangle ,
\label{eq:2.36}
\end{equation}
with
\begin{equation}
J_{\mathbf {r}} \equiv [a_{\mathbf {r}} , V] . \label{eq:2.37}
\end{equation}

The structure of $\mcm_n(z)$ is identical to that of the Feshbach's
potential, but its one-body expression is different. For instance,
the Feshbach's potential is not symmetric for the exchange $\r
\longleftrightarrow \rf$. In contrast, $\mcm_n(z)$ is symmetric, as
one can check working out the right-hand side of Eqs.~(\ref{eq:2.35})
and (\ref{eq:2.36}).

\section{Green's function approach to inclusive $(e,e')$ reactions
\label{sec.green}}

\subsection{Hadronic tensor \label{ssec:3.1} }

In order to avoid complications of minor interest in the present
context, we omit recoil corrections, use the one-photon exchange
approximation with one-body currents, and consider only spinless
point-like protons.

The inclusive cross section for the quasielastic $(e,e')$ scattering
on a nucleus of $Z+1$ protons is given by
\begin{equation}
\sigma = K (2\epsilon_{\mathrm L} R_{\mathrm L} + R_{\mathrm T}) ,
\label{eq:3.1}
\end{equation}
where $K$ is a kinematical factor and
\begin{equation}
\epsilon_{\mathrm L} = -\frac {q^2} {|\q|^2} \left( 1 - 2 \frac
{|\q|^2} {q^2} \tan^2 \frac {\theta} {2} \right)^{-1} \label{eq:3.2}
\end{equation}
measures the polarization of the virtual photon. In
Eq.~(\ref{eq:3.2}), $\theta$ is the scattering angle of the electron
and $q^2 = \omega^2 - |\q|^2$, where ($\omega,\q$) is the
four-momentum transfer. All nuclear structure information is contained
in the longitudinal and transverse response functions, defined by
\begin{eqnarray}
 R_{\mathrm L}(\omega,\q) & = & W^{00}(\omega,\q), \nonumber \\
 R_{\mathrm T}(\omega,\q) & = & W^{11}(\omega,\q) + W^{22}(\omega,\q)
\label{eq:3.3}
\end{eqnarray}
in terms of the diagonal components of the hadronic tensor
\begin{eqnarray}
W^{\mu\mu}(\omega,\q) & = & \langle \psi_0| J^{\mu\dagger}(\q)
\delta(\omega + E_0 - H) J^{\mu}(\q)|\psi_0\rangle \nonumber \\
 & = & \bint\sum_{\textrm {f}}\langle \psi_0| J^{\mu\dagger}(\q)
 |\psi_f\rangle\langle \psi_f|J^{\mu}(\q)|\psi_0\rangle
 \delta(\omega + E_0 - E_f).
\label{eq:3.4}
\end{eqnarray}
Here $J^{\mu}(\q)$ is the one-body nuclear charge-current operator,
$|\psi_0\rangle$ is the initial state of the ($Z+1$)-protons target
nucleus of energy $E_0$ and $|\psi_f\rangle$ is the corresponding
final state of energy $E_f=E_0+\omega$. The target ground state
$|\psi_0\rangle$ is assumed to be nondegenerate. The sum runs over
the target bound states and the scattering states corresponding to a
proton scattered from the residual nucleus in a bound state
$|n\rangle$ or in an unbound state $|\epsilon \rangle$. For sake of
simplicity, the degeneracy indices are suppressed. All the states are
properly antisymmetrized. Accordingly, the nuclear Hamiltonian $H$
and the current components $J^{\mu}(\q)$ are understood as second
quantization operators.

\subsection{Energy sum rules for the incoherent and the coherent
contributions \label{ssec:3.2}}

Acting on $|\psi_0\rangle$ the scalar component
\begin{equation}
J^0(\q) = \int \diff \r \, \exp (i\q\cdot\r) \, a_{\mathbf
{r}}^{\dagger} a_{\mathbf {r}} \label{eq:3.5}
\end{equation}
yields the wave function
\begin{equation}
\sum_{i=1}^{Z+1} J_i^0(\q,\r_i) \psi_0 (\r_1,...,\r_{Z+1}) \, , \,
J_i^0(\q,\r_i) = \exp (i\q\cdot\r_i). \label{eq:3.6}
\end{equation}
Similar expressions are obtained for the other components $J^\mu$
of the current operator. The hadronic tensor of Eq.~(\ref{eq:3.4}) is
the sum of the incoherent contribution
\begin{equation}
W_{\mathrm{inc}}^{\mu\mu}(\omega,\q) = (Z+1) \langle \psi_0|
J_1^{\mu\dagger}(\q)\delta(\omega + E_0 - H) J_1^{\mu}(\q)|\psi_0
\rangle \label{eq:3.7}
\end{equation}
and the coherent one
\begin{equation}
W_{\mathrm{coh}}^{\mu\mu}(\omega,\q) = (Z+1) \sum_{i=2}^{Z+1}\langle
\psi_0| J_1^{\mu\dagger}(\q)\delta(\omega + E_0 - H)
J_i^{\mu}(\q)|\psi_0 \rangle . \label{eq:3.8}
\end{equation}

The non-energy-weighted electromagnetic sum rule deals with the
integrated strength
\begin{equation}
\Sigma^{\mu\mu}(\q) = \int_{-0}^{\infty}\diff \omega \,
W^{\mu\mu}(\omega,\q) = \langle \psi_0
|J^{\mu\dagger}(\q)J^{\mu}(\q)|\psi_0 \rangle , \label{eq:3.9}
\end{equation}
where the lower integration limit means that the integral includes
the $\delta$--singularity at $\omega$=0 (full sum rule inclusive of
the elastic contribution). In the scalar case one has
\begin{equation}
\Sigma_{\mathrm{inc}}^{00}(\q) = Z + 1 , \label{eq:3.10}
\end{equation}
\begin{equation}
\Sigma_{\mathrm{coh}}^{00}(\q) = (Z + 1) \sum_{i=2}^{Z+1} \langle
\psi_0 |J_1^{0\dagger}(\q)J_i^{0}(\q)|\psi_0 \rangle .
\label{eq:3.11}
\end{equation}

In the Fermi gas model $\Sigma_{\mathrm{coh}}^{00}(\q)$ is negligible
for $|\q| > 2k_{\mathrm F}$, where $k_{\mathrm F}$ is the Fermi
momentum. This is still approximately true in presence of
correlations [Refs.~\cite{fabro,jourd}]. Therefore one has
\begin{equation}
\Sigma^{00}(\q) \simeq (Z + 1) \quad \mathrm{for} \quad |\q| > 2
k_{\mathrm F} . \label{eq:3.12}
\end{equation}

\subsection{Projection operator method \label{ssec:3.3}}

The approach developed in this Section differs form previous
treatments \cite{chinn,capuzzi,capma} for three reasons.

i) We want to include the effects of the final state interaction in
terms of the self-energy, rather than in terms of the Feshbach
optical potential or of the particle self-energy. In fact, the full
self-energy is a more fundamental quantity, has no mathematical
drawbacks, and is more closely related to the empirical optical-model
potentials. Therefore, we shall not use the Feshbach's projection
operators as in Ref.~\cite{capuzzi}, but the extended ones of
Eq.~(\ref{eq:2.20}), already used in Ref.~\cite{capma}.

ii) We want to include the interference between different channels
$|n\rangle$, neglected in Ref.~\cite{capma}, which is useful to
express the hadronic tensor in terms of the local potential
equivalent to the self-energy. This requires some changes in the
approach of Ref.~\cite{capma}, which give in practice the same result
when the interference effects are disregarded.

iii) We want to introduce different approximations for the elastic
and inelastic contributions to the hadronic tensor, when dealing with
the dependence on the state of the residual nucleus.

In order to keep the treatment as simple as possible, we shall only
refer to the scalar component $W^{00}(\omega,\q)$, with $J^0(\q)$
written in the momentum representation, i.e.
\begin{equation}
J^{0}(\q) = \int \diff\p \, a_{\mathbf {p}}^{\dagger} a_{\mathbf
{p-q}} . \label{eq:3.13}
\end{equation}
Disregarding the upper indices of the hadronic tensor and inserting
the completeness relation of the residual-nucleus states, Eq.~(\ref{eq:3.4})
yields
\begin{equation}
W(\omega,\q) = \bint\sum_n \int\diff \p \, \langle \psi_0| a_{\mathbf
{p-q}}^{\dagger} |n\rangle\langle n| a_{\mathbf {p}} \ \delta(\omega
+ E_0 - H) J^{0}(\q)|\psi_0\rangle , \label{eq:3.14}
\end{equation}
where the sum is understood over $|n\rangle$ and $|\epsilon\rangle$.
Equation (\ref{eq:3.14}) was obtained in Ref.~\cite{capma} in a more
complicated way. Here, the same result is recovered with the
insertion of a completeness relation into the second quantization
expression of $J^{0}(\q)$ and has the same generality, i.e. can be
applied to every one-body charge-current operator.

As was done in previous papers, we shall now disregard the
contribution of the continuum states of the residual nucleus. Although
this approximation is correct at the energy and momentum transfers
considered here, we shall recover the continuum contribution in
Sec.~\ref{sssec:3.4.3}. Equation (\ref{eq:3.14}) is approximated with
\begin{equation}
W(\omega,\q) = \sum_n \mathrm{Re} \int\diff \p \, \langle \psi_0|
a_{\mathbf {p-q}}^{\dagger} |n\rangle\langle n| a_{\mathbf {p}} \
\delta(\omega + E_0 - H) J^{0}(\q)|\psi_0\rangle , \label{eq:3.15}
\end{equation}
where the real part has been extracted in order to restore the real
nature of $W(\omega,\q)$, which can be lost after the approximation.
This is equivalent to introduce the truncated completeness
symmetrically in $J^{0\dagger}(\q)$ and in $J^{0}(\q)$.

Now, the treatment of Ref.~\cite{capma} is slightly modified in order
to obtain expressions directly related to the self-energy, according
to the above items i) and ii). This requires the introduction of the
vectors $(a_{\mathbf {p}}^{\dagger} + a_{\mathbf {p}})|n\rangle$,
instead of $a_{\mathbf {p}}^{\dagger}|n\rangle$, and the use of
Eq.~(\ref{eq:2.24}) in order to feature $\hat H_n$. To this extent, we
add to the second factor of Eq.~(\ref{eq:3.15}), written as
\begin{equation}
\langle n| a_{\mathbf {p}} \ \delta(\omega + E_0 - \epsilon_n - \hat
H_n) J^{0}(\q)|\psi_0\rangle , \label{eq:3.15ab}
\end{equation}
the term
\begin{equation}
\langle n|a_{\mathbf {p}}^{\dagger}\, \delta(\omega + E_0 -\epsilon_n
- \hat H_n)J^{0}(\q)|\psi_0\rangle = \langle n|a_{\mathbf
{p}}^{\dagger}\, \delta(\omega + E_0 - 2\epsilon_n +
H)J^{0}(\q)|\psi_0\rangle, \label{eq:3.19c}
\end{equation}
which is null, since it is a scalar product between states with a
different number of protons. Moreover, we add to the terms
(\ref{eq:3.15ab}) and (\ref{eq:3.19c}) the corresponding ones with
$\omega$ replaced by $-\omega$, i.e.
\begin{equation}
\langle n|a_{\mathbf {p}}\, \delta(-\omega + E_0 - \epsilon_n - \hat
H_n) J^{0}(\q)|\psi_0\rangle = \langle n|a_{\mathbf {p}}\,
\delta(-\omega + E_0 - H) J^{0}(\q)|\psi_0\rangle, \label{eq:3.19b}
\end{equation}
which is null for $\omega > 0$ because the $\delta$-function is not
fed with the eigenfunctions of $H$, and
\begin{eqnarray}
\langle n|a_{\mathbf {p}}^{\dagger}\, \delta(-\omega + E_0 -
\epsilon_n &-& \hat H_n)J^{0}(\q)  |\psi_0\rangle = \nonumber\\
\langle n| a_{\mathbf{p}}^{\dagger}\, \delta(&-&\omega + E_0 -
2\epsilon_n + H)J^{0}(\q)|\psi_0\rangle, \label{eq:3.19a}
\end{eqnarray}
which is null for the same reason as the term (\ref{eq:3.19c}). The
terms (\ref{eq:3.19b}) and (\ref{eq:3.19a}) are
necessary to obtain a null sum rule for the interference
contribution, as will be apparent below. We emphasize that the term 
(\ref{eq:3.19b}) can be added only for $\omega \neq 0$.

After addition of the null terms (\ref{eq:3.19c})--(\ref{eq:3.19a}),
Eq.~(\ref{eq:3.15}) reads
\begin{eqnarray}
W(\omega,\q) = A(\omega,\q) + A(-\omega,\q) \label{eq:3.16}
\end{eqnarray}
with
\begin{eqnarray}
 A(\omega,\q) &=& \sum_n \mathrm{Re} \int\diff \p \, \langle \psi_0|
a_{\mathbf {p-q}}^{\dagger} |n\rangle \nonumber \\
& \times & \langle n| (a_{\mathbf {p}} +
a_{\mathbf {p}}^{\dagger})\ \delta(\omega + E_0 - \epsilon_n - \hat
H_n) J^{0}(\q)|\psi_0\rangle . \label{eq:3.17}
\end{eqnarray}
In Eq.~(\ref{eq:3.16}), as well as in the following analogous
expressions, it is understood that the $\delta$--singularity at
$\omega= 0$ must be retained only in $A(\omega,\q)$ to avoid a double
counting. Therefore, one has
\begin{equation}
\int_{-0}^{\infty} \diff\omega \, W(\omega,\q) =
\int_{-\infty}^{\infty} \diff\omega \, A(\omega,\q). \label{eq:3.18}
\end{equation}

Now we use the extended projection operators, defined in
Eq.~(\ref{eq:2.20}),
\begin{equation}
P_n =\int \diff\p \, \alpha_{\mathbf {p}}|n\rangle\langle
n|\alpha_{\mathbf {p}} \quad \mathrm{and}\quad Q_n = 1 - P_n
\label{eq:3.20}
\end{equation}
to separate in Eq.~(\ref{eq:3.17}) a direct ($A^{\mathrm D}$) and an
interference ($A^{\mathrm I}$) term, such that
\begin{equation}
A(\omega,\q) = A^{\mathrm D}(\omega,\q) + A^{\mathrm I}(\omega,\q)
\label{eq:3.21}
\end{equation}
with
\begin{eqnarray}
A^{\mathrm D}(\omega,\q) &=& \sum_n \mathrm{Re} \int\diff \p \,
\langle \psi_0| a_{\mathbf {p-q}}^{\dagger} |n\rangle  \nonumber \\
& \times & \langle n|(\alpha_{\mathbf {p}} P_n \, \delta(\omega + E_0
-\epsilon_n -\hat H_n) P_n J^{0}(\q)|\psi_0\rangle , \label{eq:3.22a}
\end{eqnarray}
\begin{eqnarray}
A^{\mathrm I}(\omega,\q) &=& \sum_n \mathrm{Re} \int\diff \p \,
\langle \psi_0| a_{\mathbf{p-q}}^{\dagger} |n\rangle  \nonumber \\
& \times & \langle n|(\alpha_{\mathbf {p}} P_n \, \delta(\omega + E_0
-\epsilon_n -\hat H_n) Q_n J^{0}(\q)|\psi_0\rangle . \label{eq:3.22b}
\end{eqnarray}
Accordingly, the hadronic tensor $W(\omega,\q)$ is decomposed into a
direct and an interference contribution:
\begin{equation}
W^{\mathrm D}(\omega,\q) = A^{\mathrm D}(\omega,\q) + A^{\mathrm
D}(-\omega,\q) \label{eq:3.23a}
\end{equation}
\begin{equation}
W^{\mathrm I}(\omega,\q) = A^{\mathrm I}(\omega,\q) + A^{\mathrm
I}(-\omega,\q) \label{eq:3.23b}
\end{equation}
We remark that $W^{\mathrm I}$ does not contribute to the sum rule,
due to the relation $P_nQ_n=0$ and to the relation
\begin{equation}
\int_{-\infty}^{\infty} \diff\omega \, \delta(\omega + E_0 -
\epsilon_n - \hat H_n) = \int_{-\infty}^{\infty} \diff E \, \delta(E
- \hat H_n) = 1 , \label{eq:3.24}
\end{equation}
which yields
\begin{equation}
\int_{-0}^{\infty} \diff\omega \, W^{\mathrm I}(\omega,\q) =
\int_{-\infty}^{\infty} \diff \omega \, A ^{\mathrm I}(\omega,\q) = 0
. \label{eq:3.25}
\end{equation}

Note that the terms (\ref{eq:3.19b}) and (\ref{eq:3.19a}), which yield
the contribution of the negative energies to the second integral, are
essential to fulfill Eq.~(\ref{eq:3.25}).

We emphasize that the insertion of the terms
(\ref{eq:3.19c})--(\ref{eq:3.19a}) does not change the expression of
$W(\omega,\q)$, but produces an effect on its decomposition into
$W^{\mathrm D}(\omega,\q)$ and $W^{\mathrm I}(\omega,\q)$. As a
consequence, the present decomposition is slightly different form
that of Ref.~\cite{capma}. The term~(\ref{eq:3.19c}), absent in
Ref.~\cite{capma}, modifies the decomposition at very low energies,
i.e. at $\omega \leq 2\epsilon_n - E_0 - \epsilon_0^{A-1}$, where
$\epsilon_0^{A-1}$ is the ground state energy of the ($Z-1$)--body
Hamiltonian. The term~(\ref{eq:3.19b}) produces no effects. The term
(\ref{eq:3.19a}), which is also present in Ref.~\cite{capma}, differs
here by a shift of $2(E_o - \epsilon_n)$ in the value of the
argument. This shift is the only difference from the treatment of
Ref.~\cite{capma} at the energies $\omega$ considered in this paper.
The changes introduced here influence only the separation between the
direct and the interference contributions, allowing a more
appropriate treatment of the latter, which in Ref.~\cite{capma} is
totally neglected. The effects of these changes on $W^{\mathrm
D}(\omega,\q)$ are in practice negligible at high momentum transfers, 
as it will be seen in Sec.~\ref{sssec:3.4.5}.

\subsection{Direct contribution to the hadronic tensor
\label{ssec:3.4}}

\subsubsection{One-body expression of the incoherent and coherent
contributions \label{sssec:3.4.1}}

Using (\ref{eq:3.20}), Eq.~(\ref{eq:3.22a}) is expressed in terms of
vectors and operators of a s.p. Hilbert space as
\begin{equation}
A^{\mathrm D}(\omega,\q) = \sum_n \mathrm{Re} \langle \phi_n|
j^{0\dagger}(\q) \mcs_n(\omega_n)|\Phi_n(\q)\rangle \, , \, \omega_n
= \omega + E_0 - \epsilon_n, \label{eq:3.26}
\end{equation}
where $|\phi_n\rangle$ and $\mcs_n$ represent the hole overlaps and
the one-body spectral functions, defined by Eqs.~(\ref{eq:2.1}) and
(\ref{eq:2.28a}), respectively, and
\begin{equation}
\langle \p| j^{0}(\q)|\phi_n\rangle = \langle \p - \q|\phi_n\rangle =
\langle n|a_ {\mathbf {p-q}}|\psi_0\rangle , \label{eq:3.27}
\end{equation}
\begin{equation}
 \langle \p| \Phi_n(\q)\rangle = \langle n|a_ {\mathbf {p}}
 J^{0}(\q)|\psi_0\rangle .
\label{eq:3.28}
\end{equation}

Using the relation
\begin{equation}
[a_{\mathbf {p}},J^{0}(\q)] = a_ {\mathbf {p-q}} \label{eq:3.29}
\end{equation}
in Eq.~(\ref{eq:3.26}), the contribution to $|\Phi_n(\q)\rangle$ due
to the many-body current $J^{0}(\q)$ is split into the contribution of
a single proton and that of all the other residual $Z$ protons, as
\begin{equation}
\langle \p| \Phi_n(\q)\rangle = \langle \p | j^{0}(\q)|\phi_n\rangle
+ \langle \p |\delta \Phi_n(\q)\rangle , \label{eq:3.30}
\end{equation}
where
\begin{eqnarray}
& & \langle \p |\delta \Phi_n(\q)\rangle \equiv \langle n
|J^{0}(\q)
a_ {\mathbf {p}}|\psi_0\rangle \nonumber \\
 = \sqrt {Z+1} & \int & \diff \p_2 ... \diff \p_{Z+1}
\langle n|\p_2 ... \p_{Z+1}\rangle\langle\p,\p_2 ... \p_{Z+1}|
\sum_{i=2}^{Z+1} J_i^{0}(\q)|\psi_0 \rangle . \label{eq:3.31}
\end{eqnarray}
From Eq.~(\ref{eq:3.30}), we express $W^{\mathrm D}(\omega,\q)$ as the
sum of the terms
\begin{eqnarray}
W_{\mathrm{inc}}^{\mathrm D}(\omega,\q) & = &
A_{\mathrm{inc}}^{\mathrm D}(\omega,\q) + A_{\mathrm{inc}}^{\mathrm
D} (-\omega,\q), \nonumber \\
A_{\mathrm{inc}}^{\mathrm D}(\omega,\q) & = & \sum_n \mathrm{Re}
\langle \phi_n |j^{0\dagger}(\q) \mcs_n (\omega_n) j^{0}(\q)|
\phi_n\rangle \label{eq:3.32}
\end{eqnarray}
and
\begin{eqnarray}
W_{\mathrm{coh}}^{\mathrm D}(\omega,\q) & = &
A_{\mathrm{coh}}^{\mathrm D}(\omega,\q) + A_{\mathrm{coh}}^{\mathrm
D} , (-\omega,\q), \nonumber \\
A_{\mathrm{coh}}^{\mathrm D}(\omega,\q) & = & \sum_n \mathrm{Re}
\langle \phi_n |j^{0\dagger}(\q)
\mcs_n(\omega_n)|\delta\Phi_n(\q)\rangle . \label{eq:3.33}
\end{eqnarray}
The decomposition (\ref{eq:3.30}) is exactly equivalent to that of
Eqs.~(\ref{eq:3.7}) and (\ref{eq:3.8}). In keeping with this
correspondence, we call $W_{\mathrm{inc}}^{\mathrm D}$ and
$W_{\mathrm{coh}}^{\mathrm D}$ the {\lq\lq incoherent\rq\rq} and
{\lq\lq coherent\rq\rq} contributions to $W^{\mathrm D}$.

\subsubsection{Dependence on the state of the residual nucleus}

No information is available concerning the spectral functions
$\mcs_n(E)$ related to the excited states of the residual nucleus.
Therefore, we are obliged to express $\mcs_n(E)$ in terms of
$\mcs_0(E)$. According to previous papers
\cite{fesh3,moniz,moniz2}, we assume: i) $\mcs_n(E)$ only
differs from $\mcs_0(E)$ by an energy shift $\delta\epsilon_n$, ii)
the contributions of the various shifts can be taken into account by
means of a single average shift $\tilde \epsilon$. Thus, in
Eqs.~(\ref{eq:3.26}), (\ref{eq:3.32}), and (\ref{eq:3.33}) we set
\begin{equation}
\mcs_n(\omega + E_0 - \epsilon_n) \simeq \mcs_0(\tilde\omega) \, ,
\, \tilde\omega = \omega + E_0 - \epsilon_0 - \tilde\epsilon .
\label{eq:3.34}
\end{equation}
The choice of $\tilde\epsilon$ will be discussed later.

\subsubsection{Inclusion of the unbound states of the residual
nucleus \label{sssec:3.4.3} }

For a single state $|\epsilon\rangle$ of the continuous spectrum, one
can define neither an extended projection operator nor a Feshbach's
one. This is the reason why the states $|\epsilon\rangle$ are usually
neglected in the treatments based on the Green's function approach.
Even if the continuum contribution is negligible at the energy and
momentum transfers considered here, the lack of completeness
resulting if one considers only the bound states $|n\rangle$ appears
unsatisfactory from a conceptual point of view, mainly because the
sum rule is not exactly fulfilled. This drawback can be easily
eliminated if Eq.~(\ref{eq:3.34}) is used. In fact, from
Eq.~(\ref{eq:3.26}), by expliciting the scalar products and using
Eqs.~(\ref{eq:3.27}) and (\ref{eq:3.28}), we have
\begin{eqnarray}
A^{\mathrm D}(\omega,\q) & = & \sum_n \mathrm{Re} \int \diff\p
\diff\p' \langle \psi_0 |a_{\mathbf {p-q}}^{\dagger}|n\rangle \langle
n|a_{\mathbf {p'}} J^{0}(\q) |\psi_0\rangle \langle \p|
\mcs_0(\tilde\omega)|\p'\rangle , \nonumber \\ \tilde\omega & = &
\omega + E_0 -\epsilon_0 - \tilde\epsilon . \label{eq:3.35}
\end{eqnarray}

Here, it is quite natural to recover the contribution of the
continuous spectrum after making the substitution
\begin{equation}
\sum_n |n \rangle \langle n| \longrightarrow \sum_n |n \rangle
\langle n| + \int \diff\epsilon \, |\epsilon\rangle \langle\epsilon|
= 1 \label{eq:3.36}
\end{equation}
which yields
\begin{eqnarray}
A^{\mathrm D}(\omega,\q) & = & \mathrm{Re} \int \diff\p \diff\p'
\langle \psi_0 |a_{\mathbf {p-q}}^{\dagger} a_{\mathbf {p'}}
J^{0}(\q) |\psi_0\rangle \langle \p| \mcs_0(\tilde\omega)|\p'\rangle
\nonumber \\ & = & \mathrm{Re}\,\mathrm{Tr} [\, {\overline
K}_{\mathbf {q}} \mcs_0(\tilde\omega)], \label{eq:3.37}
\end{eqnarray}
where
\begin{eqnarray}
\langle \p'|\,{\overline K}_{\mathbf {q}} |\p\rangle \equiv \langle
\psi_0 |a_{\mathbf {p-q}}^{\dagger} a_{\mathbf {p'}} J^{0}(\q)
|\psi_0\rangle
%\nonumber \\
 = \int \diff \k \,\langle \psi_0 |a_{\mathbf {p-q}}^{\dagger}
a_{\mathbf {p'}}a_{\mathbf {k}}^{\dagger} a_{\mathbf
{k-q}}|\psi_0\rangle . \label{eq:3.38}
\end{eqnarray}
Analogously, using Eq.~(\ref{eq:3.29}), one has
\begin{equation}
A_{\mathrm {inc}}^{\mathrm D}(\omega,\q) = \mathrm{Re}\,\mathrm{Tr}
[K_{\mathbf {q}} \mcs_0(\tilde\omega)] \label{eq:3.39}
\end{equation}
with
\begin{equation}
\langle \p'| K_{\mathbf {q}} |\p\rangle = \langle \psi_0 | a_{\mathbf
{p-q}}^{\dagger} a_{\mathbf {p'-q}}|\psi_0\rangle \label{eq:3.40}
\end{equation}
and
\begin{equation}
A_{\mathrm {coh}}^{\mathrm D}(\omega,\q) = \mathrm{Re}\,\mathrm{Tr}
[\Delta K_{\mathbf {q}} \mcs_0(\tilde\omega)] \label{eq:3.41}
\end{equation}
with
\begin{equation}
\langle \p'| \Delta K_{\mathbf {q}} |\p\rangle = \int \diff \k \,
\langle \psi_0 | a_{\mathbf {p-q}}^{\dagger} a_{\mathbf
{k}}^{\dagger}a_{\mathbf {k-q}} a_{\mathbf {p'}}|\psi_0\rangle .
\label{eq:3.42}
\end{equation}

Since $K_{\mathbf {q}}$ and $\mcs_0$ are nonnegative (hermitian)
operators and $K_{\mathbf {q}}$ is self-adjoint, $\mathrm{Tr}
[K_{\mathbf {q}} \mcs_0(\tilde\omega)]$ is nonnegative, as it follows
easily by expressing the trace on the basis of the eigenvectors of
$K_{\mathbf {q}}$. Thus one has
\begin{equation}
A_{\mathrm {inc}}^{\mathrm D}(\omega,\q) = {\mathrm {Tr}} 
\, [K_{\mathbf {q}} \mcs_0( \tilde \omega )] \ge 0 \, . \label{eq:3.45}
\end{equation}

The expression of
\begin{equation}
W^{\mathrm D}(\omega,\q) = A^{\mathrm D}(\omega,\q) + A^{\mathrm
D}(-\omega,\q) ,
\end{equation}
obtained from Eq.~(\ref{eq:3.37}), and the corresponding expressions
of $W_{\mathrm{inc}}^{\mathrm D}(\omega,\q)$ and
$W_{\mathrm{coh}}^{\mathrm D}(\omega,\q)$, are essentially the same
as those obtained in Secs.~9.1.1 and 9.1.3 of Ref.~\cite{capma}. This
can be easily checked noting that the spectral function $\mcs_0(E)$ is
equal to the particle (hole) spectral function at positive (negative)
energies. The only differences from the results of Ref.~\cite{capma}
are the insertion of the average energy $\tilde \epsilon$ and an
energy shift in $A_{\mathrm {inc}}^{\mathrm D}(-\omega,\q)$ and in
$A_{\mathrm {coh}}^{\mathrm D}(-\omega,\q)$.

The problem of defining a projection operator related to a single
continuous eigenvalue $\epsilon$ of the residual nucleus is of
strictly mathematical nature. It is solved in the Appendix
associating with $\epsilon$ a set of approximate eigenvectors
depending on an index $\eta$ which describes the accuracy of the
approximation (increasing for $\eta \rightarrow +0$). They can
replace the exact eigenvectors $|\epsilon \rangle$ in the whole
formalism for two reasons. i) In the limit for $\eta \rightarrow +0$
they satisfy a completeness relation analogous to Eq.~(\ref{eq:3.36}).
ii) One can associate with $\epsilon$ a projection operator since the
approximate eigenvectors are normalizable.

\subsubsection{Non-energy-weighted sum rules}

We shall prove here that the approximation of Eq.~(\ref{eq:3.34}),
relating the spectral functions $\mcs_n(E)$ to $\mcs_0(E)$, has no
effect on the sum rules, both for the incoherent and the coherent
contributions to the hadronic tensor. Since in Eq.~(\ref{eq:3.25}) we
proved that the interference term $W^{\mathrm I}(\omega,\q)$ does not
contribute to the sum rule, Eqs.~(\ref{eq:3.37}), (\ref{eq:3.39}), and
(\ref{eq:3.41}) should satisfy the same sum rules as the exact
hadronic tensor.

Remembering Eq.~(\ref{eq:3.10}) and using Eq.~(\ref{eq:2.13}), i.e.
\begin{equation}
\int_{-\infty}^{+\infty} \diff E \, \mcs_0(E) = 1 \, ,
\label{eq:3.46}
\end{equation}
Eq.~(\ref{eq:3.37}) yields
\begin{eqnarray}
\Sigma^{\mathrm D}(\q) & \equiv & \int_{-0}^{\infty} \diff \omega \,
W^{\mathrm D}(\omega,\q) = \int_{-\infty}^{\infty} \diff \omega \,
A^{\mathrm D}(\omega,\q) \nonumber \\
& = & {\mathrm {Re}}\, {\mathrm {Tr}} \, \left[\, {\overline
K}_{\mathbf {q}} \int_{-\infty}^{\infty} \diff \tilde \omega \,
\mcs_0(\tilde \omega) \right] = {\mathrm {Re}}\,{\mathrm {Tr}} \,
[\,{\overline K}_{\mathbf {q}}]. \label{eq:3.47}
\end{eqnarray}
By Eqs.~(\ref{eq:3.38}) and (\ref{eq:3.13}), one has
\begin{equation}
{\mathrm {Tr}} \, [\,{\overline K}_{\mathbf {q}}] = \int\diff \p
\diff \k \, \langle \psi_0| a_{\mathbf {p-q}}^{\dagger}a_{\mathbf
{p}}a_{\mathbf {k}}^{\dagger} a_{\mathbf {k-q}}|\psi_0\rangle =
\langle \psi_0|J^{0\dagger}(\q) J^{0}(\q)|\psi_0\rangle
\label{eq:3.48}
\end{equation}
and so
\begin{equation}
\Sigma^{\mathrm D}(\q) =\langle \psi_0|J^{0\dagger}(\q)
J^{0}(\q)|\psi_0\rangle , \label{eq:3.49}
\end{equation}
according to the exact sum rule of Eq.~(\ref{eq:3.9}). Likewise, using
Eqs.~(\ref{eq:3.39}) and (\ref{eq:3.40}), one has
\begin{equation}
\Sigma_{\mathrm {inc}}^{\mathrm D}(\q) \equiv \int_{-0}^{\infty}
\diff \omega \, W_{\mathrm {inc}}^{\mathrm D}(\omega,\q) = {\mathrm
{Tr}} \, [K_{\mathbf {q}}] = {\mathrm {Tr}} \, [K] = Z + 1,
\label{eq:3.50}
\end{equation}
according to the exact sum rule of Eq.~(\ref{eq:3.10}). It follows
that also the sum rule for the coherent part is in accordance with
Eq.~(\ref{eq:3.11}).

\subsubsection{Practical approximations for the direct contribution to 
the hadronic tensor \label{sssec:3.4.5} }

We have previously expressed the direct contribution $W^{\mathrm
D}(\q)$ to the hadronic tensor as the sum of the incoherent and
coherent parts
\begin{equation}
W_{\mathrm {inc}}^{\mathrm D}(\omega,\q) = A_{\mathrm {inc}}^{\mathrm
D}(\omega,\q) + A_{\mathrm {inc}}^{\mathrm D}(-\omega,\q) ,
\label{eq:3.51}
\end{equation}
\begin{equation}
W_{\mathrm {coh}}^{\mathrm D}(\omega,\q) = A_{\mathrm {coh}}^{\mathrm
D}(\omega,\q) + A_{\mathrm {coh}}^{\mathrm D}(-\omega,\q) .
\label{eq:3.52}
\end{equation}
In this Subsection, we profit from some results of Ref.~\cite{capma}
to simplify $W^{\mathrm D}(\omega,\q)$ in view of practical
calculations. We notice that the terms of Eqs.~(\ref{eq:3.51}) and
(\ref{eq:3.52}) at positive (negative) arguments correspond to the
particle (hole) contributions of the quoted reference.

In the calculations based on the Green's function approach, the
coherent part of the hadronic tensor is usually disregarded since it
is too difficult to evaluate. Consequently, only the sum rule
$\Sigma_{\mathrm {inc}}^{\mathrm D}(\q) = Z + 1$ of (\ref{eq:3.50}) is
relevant. This is considered correct at high momentum transfers,
schematically for $q > k_{\mathrm F}$. In Sec.~10 of
Ref.~\cite{capma}, it was shown that this approximation is excessive
even for uncorrelated systems, since only $A_{\mathrm {coh}}^{\mathrm
D}(\omega,\q)$ is negligible at $q > k_{\mathrm F}$, whereas
$A_{\mathrm {coh}}^{\mathrm D}(-\omega,\q)$ is still sizable for
$k_{\mathrm F} < q < 2k_{\mathrm F}$. Fortunately, the latter term is
largely cancelled by $A_{\mathrm {inc}}^{\mathrm D}(-\omega,\q)$ and
this produces the simplification
\begin{equation}
W^{\mathrm D}(\omega,\q) = A_{\mathrm {inc}}^{\mathrm D}(\omega,\q)
\quad \mathrm{at} \quad q > k_{\mathrm F}. \label{eq:3.53}
\end{equation}

The energy shifts that are introduced here in the arguments of $A_{\mathrm
{inc}}^{\mathrm D}(-\omega,\q)$ and $A_{\mathrm {coh}}^{\mathrm
D}(-\omega,\q)$ do not influence the cancellation found in
Ref.~\cite{capma}.

Due to this cancellation, the sum rule for $W^{\mathrm D}(\omega,\q)$
is deprived of the contribution of $A_{\mathrm {inc}}^{\mathrm
D}(-\omega,\q)$, which is positive as shown in Eq.~(\ref{eq:3.45}).
Thus, the sum rule for the total hadronic tensor, which coincides
with that of $W^{\mathrm D}(\omega,\q)$ since the interference term
does not contribute, reduces to
\begin{equation}
\Sigma (\q) < Z+1 \quad \mathrm{for} \quad k_{\mathrm F} < q <
2k_{\mathrm F}. \label{eq:3.54}
\end{equation}

\subsection{Inclusion of the interference term \label{ssec:3.5}}

The interference term $W^{\mathrm I}(\omega,\q)$, defined in
Eqs.~(\ref{eq:3.22b}) and (\ref{eq:3.23b}), gives no contribution in
absence of final state correlations, since in this case $\delta(E -
\hat H_n)$ does not connect the projection operators $P_n$ and $Q_n$.
In the Green's function approach $W^{\mathrm I}(\omega,\q)$ is
usually disregarded also in presence of final-state correlations,
since it does not seem reducible to a one-body expression. As already
noticed, this approximation does not affect the sum rule. We give
here an approximated expression of $W^{\mathrm I}(\omega,\q)$, which
is reducible to a one-body expression and does not modify the sum
rule. This result has
a conceptual relevance and will be useful to introduce the empirical
optical-model potential.

Here, we are only interested in the interference term $A^{\mathrm
I}(\omega,\q)$ to be added to Eq.~(\ref{eq:3.53}). Therefore, we start
from the expression (\ref{eq:3.23b}) of $W^{\mathrm I}(\omega,\q)$ and
disregard the contribution $A^{\mathrm I}(-\omega,\q)$. Then, we
introduce an approximation which allows the reduction of $A^{\mathrm
I}(\omega,\q)$ to a one-body expression and, finally, we retain only
its incoherent part.

We insert into the expression (\ref{eq:3.22b}) of $A^{\mathrm
I}(\omega,\q)$ the Eq.~(\ref{eq:2.25}), i.e.,
\begin{equation}
\delta(E - \hat H_n) = \frac {1} {2 \pi i} [\hat G_n^{\dagger}(E) -
\hat G_n(E)] , \label{eq:3.55}
\end{equation}
with
\begin{equation}
 \hat G_n(E) \equiv \frac {1} {E - \hat H_n + i \eta} .
\label{eq:3.56}
\end{equation}
Then, we use the approximated relations
\begin{equation}
 P_n \hat G_n(E) Q_n J^0(\q)|\psi_0\rangle = - P_n \hat G_n(E)P_n
M'_n(E) J^0(\q)|\psi_0\rangle,
\label{eq:3.57}
\end{equation}
\begin{equation}
 P_n \hat G_n^{\dagger}(E) Q_n J^0(\q)|\psi_0\rangle = - P_n \hat
 G_n^{\dagger}(E)P_n (M'_n)^{\dagger}(E) J^0(\q)|\psi_0\rangle,
\label{eq:3.58}
\end{equation}
where $M'_n(E)$ is the energy derivative of the many-body self-energy
of Eq.~(\ref{eq:2.31}), obtaining
\begin{eqnarray}
A^{\mathrm I}(\omega,\q) & =& - \frac {1} {\pi} \sum_n \mathrm{Re}
\int \diff \p \, \langle \psi_0 |a_{\mathbf {p-q}}^{\dagger}|n
\rangle \nonumber \\
& \times & \langle n| \alpha_{\mathbf {p}}\frac {[\hat G_n^{\dagger}
(M'_n)^{\dagger}] (\omega_n) - (\hat G_n M'_n)(\omega_n)} {2i} J^0
(\q)| \psi_0\rangle \label{eq:3.59}
\end{eqnarray}
with $\omega_n = \omega + E_0 - \epsilon_n$.

Equations (\ref{eq:3.57}) and (\ref{eq:3.58}) have the same structure
as Eq.~(59) of Ref.~\cite{Cap} and can be deduced with the same method
replacing the Hamiltonian $H$ by $\hat H_n$ and the Feshbach's
projection operators by the extended ones. As remarked in the quoted
reference, the Eqs.~(\ref{eq:3.57}) and (\ref{eq:3.58}) must be used
inside the matrix elements of Eq.~(\ref{eq:3.59}) and hold in the
region of the quasielastic peak at intermediate and high energies.

Operating as in Sec.~\ref{sssec:3.4.1} and retaining only the
incoherent part of Eq.~(\ref{eq:3.59}), we obtain 
\begin{eqnarray}
 A_{\mathrm{inc}}^{\mathrm{I}}(\omega,\q) = - \frac {1} {\pi}
 \sum_n & \mathrm{Re} &
\langle \phi_n |j^{0\dagger}(\q) \frac {[\mcg_n^{\dagger}
(\mcm'_n)^{\dagger}]
(\omega_n) - (\mcg_n \mcm'_n)(\omega_n)} {2i} \nonumber \\
& \times & j^0 (\q)|\phi_n\rangle, \label{eq:3.60}
\end{eqnarray}
where $\mcm_n(E)$ is the one-body expression of the self-energy.
Using the relation
\begin{equation}
 \mcg_n(E)= \int_{-\infty}^{+\infty} \diff E' \, \frac {\mcs_n(E')}
 {E- E' + i\eta} ,
\label{eq:3.61}
\end{equation}
deduced from Eq.~(\ref{eq:2.11}), and the relation [see Eq.~(\ref{eq:2.17a})]
\begin{equation}
 \mcm_n(E) = E- [\mcg_n(E)]^{-1} - \mct ,
\label{eq:3.62}
\end{equation}
Eq.~(\ref{eq:3.34}) is extended to the relations
\begin{eqnarray}
&\mcg_n(\omega+E_0-\epsilon_n) \simeq \mcg_0(\tilde \omega) ,
\,\,\,\,\,\, \mcm_n(\omega+E_0-\epsilon_n) \simeq \mcm_0(\tilde
\omega)\, , \nonumber \\
& \tilde \omega = \omega + E_0 - \epsilon_0 - \tilde \epsilon .
\label{eq:3.63}
\end{eqnarray}
Therefore, the contribution of the continuous spectrum can be included
as in Sec.~\ref{sssec:3.4.3} to yield
\begin{eqnarray}
 A_{\mathrm {inc}}^{\mathrm I}(\omega,\q) = - \frac {1}{\pi}
\mathrm{Re} \, \mathrm{Tr} \,\left[ K_{\mathbf {q}} \left( \frac
{[\mcg_0^{\dagger} (\mcm'_0)^{\dagger}] (\tilde \omega) - (\mcg_0
\mcm'_0)(\tilde \omega)} {2i} \right) \right] . \label{eq:3.64}
\end{eqnarray}
Making explicit the real part of Eq.~(\ref{eq:3.64}), we have
\begin{eqnarray}
A_\mathrm{inc}^\mathrm{I}(\omega,\q) &=& - \frac{1}{2\pi}
\mathrm{Tr} \, \left[  K_{\mathbf{q}}  \left(  \frac
{[\mcg_0^{\dagger} (\mcm'_0)^{\dagger} +
(\mcm'_0)^{\dagger}\mcg_0^{\dagger}] (\tilde \omega)}
{2i} \right. \right. \nonumber \\
& - & \left. \left. \frac {(\mcg_0 \mcm'_0 + \mcm'_0 \mcg_0)(\tilde
\omega)} {2i} \right) \right] . \label{eq:3.65}
\end{eqnarray}
where $\mcm'_0(E)$ and $\mcg_0(E)$ are symmetrically arranged.

Equation (\ref{eq:3.64}) implies
\begin{equation}
\int_{-\infty}^{+\infty} \diff \omega \, A_{\mathrm {inc}}^{\mathrm
I} (\omega,\q) = 0 , \label{eq:3.66}
\end{equation}
in keeping with Eq.~(\ref{eq:3.25}). In fact, denoting by $C_{\infty}$
the large circle with center in the origin, one has
\begin{equation}
\int_{-\infty}^{+\infty} \diff E\, [\mcg_0^{\dagger} (E)
(\mcm'_0)^{\dagger}(E) - \mcg_0(E) \mcm'_0(E)] = \int_{C_{\infty}}
 \diff z \, \mcg_0(z)\mcm'_0(z) = 0 ,
\label{eq:3.67}
\end{equation}
where the first equality is due to the analyticity of
$\mcg_0(z)\mcm'_0(z)$ in the complex plane, except for cuts and poles
on the real axis, and the latter one follows from the fast decrease
of $\mcg_0(z)\mcm'_0(z)$ at infinity (more than $1/|z|$).

\subsection{Practical approximation for the total hadronic tensor
\label{ssec:3.6}}

In this Subsection we consider the total hadronic tensor as obtained
by the sum of the direct contribution of Eq.~(\ref{eq:3.53}) with the
interference term of Eq.~(\ref{eq:3.65}):
\begin{equation}
W(\omega,\q) = A_{\mathrm {inc}}^{\mathrm D}(\omega,\q) + A_{\mathrm
{inc}}^{\mathrm I}(\omega,\q) . \label{eq:3.68}
\end{equation}
Using in Eq.~(\ref{eq:3.45}), the relation (see Eq.~(\ref{eq:2.12}))
\begin{equation}
\mcs_0(E) = \frac {1} {2\pi i} [\mcg_0^{\dagger}(E) - \mcg_0(E)] ,
\label{eq:3.69}
\end{equation}
one has
\begin{equation}
W(\omega,\q) = \frac {1} {2\pi i} \mathrm{Tr} \, \left[ K_{\mathbf q}
\left( \mcg_{\mathrm {eff}}^{\dagger}(\tilde \omega) - \mcg_{\mathrm
{eff}}(\tilde \omega)\right) \right] \, , \, \tilde \omega = \omega +
E_0 - \epsilon_0 - \tilde \epsilon , \label{eq:3.70}
\end{equation}
where
\begin{equation}
\mcg_{\mathrm {eff}}(E) = [1 - \mcm'_0(E)]^{1/2} \, \mcg_0(E) [1 -
\mcm'_0(E)]^{1/2} . \label{eq:3.71}
\end{equation}
We call $\mcg_{\mathrm {eff}}(E)$ \lq\lq effective Greeen's
function\rq\rq.

Strictly, Eq.~(\ref{eq:3.71}) should involve in a symmetrical way the
factor $1 - \mcm'_0(E)/2$, instead of the square root operator. The
replacement by the latter has been done to feature the more
fundamental quantity $1 - \mcm'_0(E)$, related to the effective
$\omega$--mass (see Ref.~\cite{mabor}), and is justified by the fact
that $\mcm'_0(E)$ is small in the energy region of interest. For
instance, in the scattering $p-^{40}$Ca the approximation is very
good at proton energies beyond 50 MeV (see Fig. 2 of
Ref.~\cite{tornow}, where the result has been obtained using the
dispersion relation specific of the self-energy). Concerning the
precise meaning of the square root operator of Eq.~(\ref{eq:3.71}), we
remark that it is trivial if $\mcm'_0(E)$ is local (as supposed in
many models) and that in any case the square root can be defined
without problems as a power series for $\|\mcm'_0(E)\| \leq 1$.

The effective Green's function shows an interesting property. While
the energy derivative of $\mcg_0(E)$ satisfies the relation
\begin{equation}
\mcg'_0(E) = - \mcg_0(E) [1 - \mcm'_0(E)] \mcg_0(E), \label{eq:3.72}
\end{equation}
the derivative of $\mcg_{\mathrm {eff}}(E)$, performed with the help
of the previous equation and disregarding the contribution of
$\mcm''_0(E)$, as allowed by the quasilinear behavior of $\mcm_0(E)$
(see again Fig. 2 of Ref.~\cite{tornow}), yields
\begin{equation}
\mcg'_{\mathrm {eff}}(E) \simeq [1 - \mcm'_0(E)]^{1/2}\, \mcg'_0(E)
[1 - \mcm'_0(E)]^{1/2}= - \mcg_{\mathrm {eff}}^2(E) .
\label{eq:3.73}
\end{equation}
This is the typical relation satisfied by the Green's function of an
energy independent Hamiltonian. This property really holds. In fact
$\mcg_{\mathrm {eff}}(E)$ is invertible, since $[1 -
\mcm'_0(E)]^{-1/2}$ can be defined as a power series and, therefore,
one can define the related self-energy $\mcm_{\mathrm {eff}}(E)$ and
the Hamiltonian $h_{\mathrm {eff}}(E)$ as in Eq.~(\ref{eq:2.17a}):
\begin{equation}
\mcm_{\mathrm {eff}}(E) = h_{\mathrm {eff}}(E) - \mct \, , \,
 h_{\mathrm {eff}}(E) =E - \mcg_{\mathrm {eff}}^{-1}(E) .
\label{eq:3.74}
\end{equation}
Note that the energy derivative
\begin{equation}
\mcm'_{\mathrm {eff}}(E) = 1 + \mcg_{\mathrm {eff}}^{-1}(E)
\mcg'_{\mathrm {eff}}(E) \mcg_{\mathrm {eff}}^{-1}(E) \label{eq:3.75}
\end{equation}
is approximately equal to zero due to Eq.~(\ref{eq:3.73}). This means
that the hadronic tensor can be reduced to a one-body expression by
means of the Green's function associated to an effective self-energy,
nearly energy independent.

Equation (\ref{eq:3.70}) holds under the same conditions as
Eqs.~(\ref{eq:3.53}) and (\ref{eq:3.65}), i.e. at intermediate and
high energies, near the quasielastic peak and for $q > k_{\mathrm
F}$. Now, we arrange Eq.~(\ref{eq:3.70}) in order to feature the role
of the scalar operator $j^0(\q;\r) = \exp(i \q \cdot \r)$. This is
included in $K_{\mathbf q}$ as
\begin{equation}
\langle \r|K_{\mathbf q} |\rf\rangle = \exp(-i \q \cdot \r) K(\r,\rf)
\exp(i \q \cdot \rf) = \langle \r|j^{0\dagger}(\q)Kj^0(\q)|\rf\rangle
. \label{eq:3.76}
\end{equation}
Permuting the factors under the trace symbol and using the property
\begin{equation}
\mathrm{Tr}\,[O^{\dagger}] = {\overline{\mathrm{Tr}\,[O]}} ,
\end{equation}
Eq.~(\ref{eq:3.70}) can equivalently be written as
\begin{eqnarray}
&W(\omega,\q) = \mathrm{Tr} \, [Kj^{0\dagger}(\q)\mcs_{\mathrm
{eff}} (\tilde\omega) j^0(\q)]
 = -\frac {1} {\pi} \mathrm{Im}\, \mathrm{Tr}\, [Kj^{0\dagger}(\q)
\mcg_{\mathrm {eff}}(\tilde\omega)j^0(\q)] \, ,  \nonumber \\
& \tilde\omega = \omega + E_0 -\epsilon_0 - \tilde\epsilon,
\label{eq:3.77}
\end{eqnarray}
where
\begin{equation}
\mcs_{\mathrm {eff}}(\tilde\omega) = \frac {1} {2\pi i}
[\mcg_{\mathrm {eff}}^{\dagger}(\tilde\omega) - \mcg_{\mathrm
{eff}}(\tilde\omega)] . \label{eq:3.78}
\end{equation}

So far, we have considered only the scalar component $J^0(\q)$ of the
current in the momentum representation. The extension to an arbitrary
one-body operator
\begin{equation}
O(\q) = \int\diff\r\diff\rf a_{\mathbf r}^{\dagger} \, O(\q;\r,\rf)
a_{\mathbf r'} \label{eq:3.79}
\end{equation}
is performed on sight. Thus, Eq.~(\ref{eq:3.77}) is generalized to all
the diagonal components of the hadronic tensor, provided that only
one-body currents are considered:
\begin{eqnarray}
& W^{\mu\mu}(\omega,\q)  =  \mathrm{Tr} \, [Kj^{\mu\dagger}(\q)
\mcs_{\mathrm {eff}}(\tilde\omega) j^{\mu}(\q)]
 = -\frac {1} {\pi} \mathrm{Im}\, \mathrm{Tr}\, [Kj^{\mu\dagger}(\q)
\mcg_{\mathrm {eff}}(\tilde\omega)j^{\mu}(\q)] ,  \nonumber \\
& \tilde\omega = \omega + E_0 -\epsilon_0 - \tilde\epsilon,
\label{eq:3.80}
\end{eqnarray}

\subsection{Energy shift prescriptions \label{ssec:3.7}}

In the literature two types of energy shifts have been proposed to
connect the spectral functions $\mcs_n(E)$.

a) Kinetic energy prescription~\cite{hori,chinn,capuzzi,capma}:
\begin{equation}
\mcs_n(\omega + E_0 -\epsilon_n) = \mcs_0(\omega + E_0 -\epsilon_0 -
\delta\epsilon_n) \, , \, \delta\epsilon_n = \epsilon_n - \epsilon_0,
\label{eq:3.81}
\end{equation}
which produces an energy shift equal to $\epsilon_n - \epsilon_0$ in
the expression of the hadronic tensor. This approximation keeps the
value of the argument of the spectral function corresponding to the
kinetic energy of the emitted proton. This explains the name.

b) Total energy prescription~\cite{capma}:
\begin{equation}
\mcs_n(\omega + E_0 -\epsilon_n) = \mcs_0(\omega + E_0 -\epsilon_0 -
\delta\epsilon_n) \, , \, \delta\epsilon_n = 0, \label{eq:3.82}
\end{equation}
which yields no energy shift in the expression of the hadronic
tensor. This approximation keeps the value of the total energy of the
system given by the proton and the residual nucleus in the state
$|n\rangle$.

The same prescriptions are applied to $\mcg_n(E)$ and $\mcm_n(E)$
according to Eqs.~(\ref{eq:3.61}), and (\ref{eq:3.62}).

The total energy prescription has the merit of being simple, since it
requires no shift. In contrast, the kinetic energy prescription leads
to an expression of the hadronic tensor which involves a convolution
of $\mcs_0(E)$ with the hole spectral function related to
$|\psi_0\rangle$~\cite{capma}.

The kinetic energy prescription is the one that has been adopted by
most authors [Refs.~\cite{hori,chinn,capuzzi}]. Moreover, it is
closely related to an expression proposed for high energy transfers
[Refs.~\cite{benhar,ciofi,benhar2,sick,benhar3,benhar4}]. However,
we decided for an intermediate choice, due to the reasons explained
below.

Using the relations (see Eqs.~(\ref{eq:2.12}) and (\ref{eq:2.17}))
\begin{equation}
\mcs_n(E) = \frac {1} {2\pi i} [ \mcg_n^{\dagger}(E) - \mcg_n(E)]\, ,
\, \mcg_n(E) = \frac {1} {E - \mct - \mcm_n(E) + i \eta} ,
\label{eq:3.83}
\end{equation}
the spectral function $\mcs_n(E)$ is decomposed as
\begin{equation}
\mcs_n(E) = \mcs_n^{\mathrm {el}}(E) + \mcs_n^{\mathrm {in}}(E)
\label{eq:3.84}
\end{equation}
with
\begin{equation}
\mcs_n^{\mathrm {el}}(E) = \eta \, \mcg_n^{\dagger}
(E+i\eta)\mcg_n(E+i\eta) , \label{eq:3.85}
\end{equation}
\begin{equation}
\mcs_n^{\mathrm {in}}(E) = - \mcg_n^{\dagger} (E) \mcm_n^{\mathrm
I}(E)\mcg_n(E) \, , \, \mcm_n^{\mathrm I} = \frac {1} {2 i}
[\mcm_n(E) - \mcm_n^{\dagger}(E)] . \label{eq:3.86}
\end{equation}
The limit for $\eta \rightarrow 0$ is understood, as usual, in
Eq.~(\ref{eq:3.85}).

The pioneering paper of Ref.~\cite{hori} has shown that for $E > 0$
$\mcs_n^{\mathrm {el}}(E)$ yields the elastic contribution, due to
the overlaps $\langle n|a_{\mathbf r}|\psi_{\mathbf{k},n}\rangle$,
where $|\psi_{\mathbf{k},n}\rangle$ describes a proton of kinetic
energy $k^2/2m = E$ scattered by the residual nucleus in the state
$|n\rangle$. The asymptotic condition of incident plane wave imposes
that $\mcs_n^{\mathrm {el}}(E)$ must be related to $\mcs_0^{\mathrm
{el}}(E)$ so as to preserve $\k$ and hence the argument $E$. This is
in favour of Eq.~(\ref{eq:3.81}) rather than Eq.~(\ref{eq:3.82}). In
contrast, $\mcs_n^{\mathrm {in}}(E)$ gives the inelastic
contribution, due to the overlaps $\langle n|a_{\mathbf
r}|\psi_{\mathbf{k},m}\rangle$, with $m \neq n$, which do not contain
the plane wave. Therefore, the requirement of preserving $\k$ is not
so stringent in this case, and prevails the exigency of conserving in
Eq.~(\ref{eq:3.86}) the size of $\mcm_n^{\mathrm I}(E)$, which is
mainly determined by the number of the inelastic channels which are
open when a proton of kinetic energy $k^2/2m$ is scattered by
$|n\rangle$. This number depends on the total energy $k^2/2m +
\epsilon _n$. Hence, Eq.~(\ref{eq:3.81}) does not correctly yield the
thresholds of the inelastic processes, and Eq.~(\ref{eq:3.82}) is
favoured to relate $\mcm_n^{\mathrm I}(E)$ to $\mcm_0^{\mathrm
I}(E)$. Since the real and the imaginary parts of the self-energy
fulfill a dispersion relation (see Sec.~5.4 of Ref.~\cite{capma2}),
the total energy prescription must be applied to $\mcm_n(E)$ and hence
to $\mcg_n(E)$ and $\mcs_n^{\mathrm {in}}(E)$.

Due to the above reasons, we adopt the kinetic energy prescription
for $\mcs_n^{\mathrm {el}}(E)$ and the total energy one for
$\mcs_n^{\mathrm {in}}(E)$, i.e.,
\begin{equation}
\mcs_n(\omega + E_0 - \epsilon_n) = \mcs_0^{\mathrm {el}}(\omega +
E_0 - \epsilon_0 - \bar \epsilon) + \mcs_0^{\mathrm {in}}(\omega +
E_0 - \epsilon_0) . \label{eq:3.87}
\end{equation}
Here, $\bar \epsilon$ is taken equal to the average excitation energy
of the residual nucleus given in terms of the energies $\epsilon_n$
and the spectroscopic factors $\lambda_n$, as
\begin{equation}
\bar \epsilon = \frac {\sum_n \lambda_n (\epsilon_n - \epsilon_0)}
{\sum_n \lambda_n} . \label{eq:3.88}
\end{equation}
According to Eq.~(\ref{eq:3.87}), the effective spectral function
$\mcs_{n,\mathrm {eff}}(E)$, defined as in Eq.~(\ref{eq:3.78}), and
which differs from $\mcs_n(E)$ by corrections involving only
$\mcm'_n(E)$, is related to the ground state effective spectral
function $\mcs_{\mathrm {eff}}(E)$ of Eq.~(\ref{eq:3.78}) as
\begin{equation}
\mcs_{n,\mathrm {eff}}(\omega + E_0 - \epsilon_n) = \mcs_{\mathrm
{eff}}^{\mathrm {el}}(\omega + E_0 - \epsilon_0 - \bar \epsilon) +
\mcs_{\mathrm {eff}}^{\mathrm {in}}(\omega + E_0 - \epsilon_0),
\label{eq:3.89}
\end{equation}
where
\begin{equation}
\mcs_{\mathrm {eff}}^{\mathrm {el}}(E) = \eta \, \mcg_{\mathrm
{eff}}^{\dagger} (E+i\eta) \mcg_{\mathrm {eff}}(E+i\eta),
\label{eq:3.90}
\end{equation}
\begin{equation}
\mcs_{\mathrm {eff}}^{\mathrm {in}}(E) = - \mcg_{\mathrm
{eff}}^{\dagger} (E) \mcm_{\mathrm {eff}}^{\mathrm I}(E)
\mcg_{\mathrm {eff}}(E) \, , \, \mcm_{\mathrm {eff}}^{\mathrm I}(E)
\equiv \frac {1} {2i} (\mcm_{\mathrm {eff}}
 - \mcm_{\mathrm {eff}}^{\dagger}) ,
\label{eq:3.91}
\end{equation}
and $\mcm_{\mathrm {eff}}(E)$ is the effective mass operator defined
in Eq.~(\ref{eq:3.74}). Therefore, Eq.~(\ref{eq:3.80}) becomes
\begin{equation}
W^{\mu\mu} ( \omega , \q) = {\mathrm {Tr}} \, \big\{ Kj^{\mu \dagger}
(\q) \big[\mcs_{\mathrm {eff}}^{\mathrm {el}} (\omega + E_0
-\epsilon_0 - \bar\epsilon) + \mcs_{\mathrm {eff}}^{\mathrm {in}}
(\omega + E_0 -\epsilon_0)\big] j^\mu (\q) \big\}. \label{eq:3.92}
\end{equation}

\subsection{Hadronic tensor in terms of the local equivalent
potential \label{ssec:3.8}}

We have expressed the hadronic tensor in terms of the self-energy,
rather than in terms of the Feshbach's potential as done in
Ref.~\cite{capuzzi}, since the latter has a quite complicated
spatial nonlocality~\cite{capma2} and its kernel is not symmetric in
the coordinate representation. These drawbacks make problematic to
find a local equivalent potential to be compared with the empirical
optical-model potential. In contrast, the self-energy is symmetric,
has a simpler nonlocal structure, and is considered the theoretical
mean field most closely related to the empirical
potential~\cite{capma2}.

Here, we want to show that the hadronic tensor can be expressed in
terms of the Perey-Buck local potential, phase equivalent to the
self-energy. To this extent, we investigate the connection between
the Green's function $\mcg_0(E)$ of the self-energy $\mcm_0(E)$ and 
that related to
its local equivalent potential. We start with a simple model of
$\mcm_0(E)$, that is widely used to calculate the self-energy by means
of dispersion relations (see Sec.~4.4 of Ref.~\cite{mabor}). It
takes into account the fact that the nonlocality in the self-energy
is gathered mainly in its statical part, which is hermitian, whereas
the non-hermitian dynamical part has a weaker nonlocal structure.
Thus, one sets
\begin{equation}
\mcm_0(E) = \mcu + \mcd(E), \label{eq:3.93}
\end{equation}
where $\mcd(E)$ is non-hermitian and local, whereas the energy
independent term $\mcu$ is hermitian and has the Frahn-Lemmer
nonlocal structure
\begin{equation}
\mcu (\r , \rf) = H(|\r - \rf|) \, U\left(\frac {1} {2}|\r + \rf|
\right). \label{eq:3.94}
\end{equation}
The precise shape of the nonlocality form factor $H$ is not relevant
for the next equations. Setting
\begin{equation}
F(\q^2) \equiv \int \diff \s \, \exp (-i\q\cdot\s) H(s) ,
\label{eq:3.95}
\end{equation}
the Perey-Buck potential~\cite{PB62,cap-lectnot} is defined by the implicit equation
\begin{equation}
\mcv_{\mathrm L}(E) = U\, F\big( E-\mcv_{\mathrm L} (E)\big) +
\mcd(E) . \label{eq:3.96}
\end{equation}

Every eigenvector $|\chi_{\mathrm L}(E)\rangle$ of $\mcv_{\mathrm L}
(E)$ is related to the corresponding eigenvector $|\chi(E)\rangle$ of
$\mcm_0(E)$ by the phase conserving relation
\begin{equation}
|\chi_{\mathrm L}(E)\rangle = f(E)|\chi(E)\rangle , \label{eq:3.97}
\end{equation}
where the inverse Perey factor $f(E)$~\cite{fied,cap-lectnot} multiplies by 
the function
\begin{equation}
f(E;r) = \big[ 1 + U(r)F'\big( E-\mcv_{\mathrm L} (E;r)\big)
\big]^{1/2} . \label{eq:3.98}
\end{equation}
Higher-order corrections~\cite{cap-lectnot}, involving surface terms,
have been disregarded in Eqs.~(\ref{eq:3.96}) and (\ref{eq:3.98}).
Analogously, the Green's functions related to $\mcv_{\mathrm L}$ and
$\mcm_0$
\begin{equation}
\mcg_{\mathrm L} (E) \equiv [E - \mct - \mcv_{\mathrm L} (E) +
i\eta]^{-1} , \label{eq:3.99}
\end{equation}
\begin{equation}
\mcg_0 (E) \equiv [E - \mct - \mcm_0 (E) + i\eta]^{-1}, \label{eq:3.100}
\end{equation}
are connected by
\begin{equation}
\mcg_{\mathrm L} (E) = f(E) \mcg_0(E) f(E) . \label{eq:3.101}
\end{equation}
The energy derivatives of the two sides of Eq.~(\ref{eq:3.96}) yield
\begin{equation}
1 - \mcm'_0 (E) = f(E) [1 - \mcv'_{\mathrm L} (E)] f(E) .
\label{eq:3.102}
\end{equation}
This relation is analogous to the relation among the effective masses
$m^*$, $m_k$, and $m_\omega$ in the infinite nuclear matter (Sec.~3.6
of \cite{mabor}).

In Eqs.~(\ref{eq:3.71}), (\ref{eq:3.74}), and (\ref{eq:3.78}), we have
defined the effective Green's function of $\mcg_0$ and the linked
Hamiltonian, self-energy, and spectral function as
\begin{eqnarray}
\mcg_{\mathrm {eff}}(E) & = & \, [1 - \mcm'_0(E)]^{1/2} \, \mcg_0 (E) [1
- \mcm'_0(E)]^{1/2} \, , \nonumber \\
 h_{\mathrm {eff}}(E) & = & \, E - \mcg_{\mathrm {eff}}^{-1}(E)\, ,
\nonumber \\
\mcm_{\mathrm {eff}}(E) & = & \, h_{\mathrm {eff}}(E) - \mct\, ,
\nonumber \\
\mcs_{\mathrm {eff}}(E) & = & \, \frac {1} {2\pi i} [\mcg_{\mathrm
{eff}}^{\dagger}(E) - \mcg_{\mathrm {eff}}(E)] . \label{eq:3.103}
\end{eqnarray}
Likewise, the corresponding quantities related to $\mcg_{\mathrm L}
(E)$ and $\mcv_{\mathrm L} (E)$ are defined as
\begin{eqnarray}
\mcg_{\mathrm {L,eff}}(E) & = & \, [1 - \mcv'_{\mathrm L}(E)]^{1/2}
\, \mcg_{\mathrm L} (E) [1 - \mcv'_{\mathrm L}(E)]^{1/2} \, ,
\nonumber
\\
 h_{\mathrm {L,eff}}(E) & = & \, E - \mcg_{\mathrm {L,eff}}^{-1}(E)\,
,
 \nonumber \\
\mcv_{\mathrm {L,eff}}(E) & = & \, h_{\mathrm {L,eff}}(E) - \mct\, ,
\nonumber \\
\mcs_{\mathrm {L,eff}}(E) & = & \, \frac {1} {2\pi i} [\mcg_{\mathrm
{L,eff}}^{\dagger}(E) - \mcg_{\mathrm {L,eff}}(E)] . \label{eq:3.104}
\end{eqnarray}
Inserting Eq.~(\ref{eq:3.102}) in the first Eq.~(\ref{eq:3.103}) and
then using Eq.~(\ref{eq:3.101}), one has
\begin{equation}
\mcg_{\mathrm {eff}} (E) = \mcg_{\mathrm {L,eff}} (E),
\label{eq:3.105}
\end{equation}
and then
\begin{equation}
h_{\mathrm {eff}} (E) = h_{\mathrm {L,eff}} (E) \, , \, \mcm_{\mathrm
{eff}} (E) = \mcv_{\mathrm {L,eff}} (E)\, , \, \mcs_{\mathrm {eff}}
(E) = \mcs_{\mathrm {L,eff}} (E) . \label{eq:3.106}
\end{equation}
We emphasize that this invariance property, which will be very useful
below, is the mere consequence of having introduced into the hadronic
tensor the contribution of the interference between different
channels.

As a consequence of Eqs.~(\ref{eq:3.105}) and (\ref{eq:3.106}), the
expression (\ref{eq:3.92}) of the hadronic tensor reads
\begin{eqnarray}
W^{\mu\mu} ( \omega , \q) = & {\mathrm{Tr}} \, \left\lbrace
Kj^{\mu \dagger} (\q) \left[  \mcs_{\mathrm {L,eff}}^{\mathrm{el}}
(\omega + E_0 -\epsilon_0 - \bar\epsilon) \right. \right. \nonumber\\
& + \left. \left. \mcs_{\mathrm{L,eff}}^{\mathrm{in}} (\omega + E_0
-\epsilon_0 ) \right]  j^\mu (\q) \right\rbrace \label{eq:3.107}
\end{eqnarray}
with
\begin{equation}
\mcs_{\mathrm {L,eff}}^{\mathrm {el}}(E) \equiv \eta \, \mcg_{\mathrm
{L,eff}}^{\dagger} (E+i\eta) \mcg_{\mathrm {L,eff}}(E+i\eta) ,
\label{eq:3.108}
\end{equation}
\begin{equation}
\mcs_{\mathrm {L,eff}}^{\mathrm {in}}(E) = - \mcg_{\mathrm
{L,eff}}^{\dagger} (E) \mcv_{\mathrm {L,eff}}^{\mathrm I}(E)
\mcg_{\mathrm {L,eff}}(E) \, , \label{eq:3.109}
\end{equation}
where $\mcv_{\mathrm {L,eff}}^{\mathrm I}(E)$ is the antihermitian
part of $\mcv_{\mathrm {L,eff}}(E)$. Using the first two
Eqs.~(\ref{eq:3.104}) and Eq.~(\ref{eq:3.99}), one readily obtains
$h_{\mathrm {L,eff}}(E)$ in terms of $h_{\mathrm {L}}(E)$:
\begin{equation}
h_{\mathrm {L,eff}}(E) = [1 - \mcv'_{\mathrm {L}}(E)]^{-1/2}
[h_{\mathrm {L}}(E) - \mcv'_{\mathrm {L}}(E) E] [1 - \mcv'_{\mathrm
{L}}(E)]^{-1/2} . \label{eq:3.111}
\end{equation}
Therefore, in principle, the hadronic tensor, as well as its elastic
and inelastic parts, can be calculated in terms of quantities related
to $\mcv_{\mathrm {L}}(E)$.

\subsection{Practical calculations \label{ssec:3.9}}

The local potential $\mcv_{\mathrm {L}}(E)$, phase equivalent to the
self-energy, can be identified with the empirical optical-model
potential. Therefore, the symbol $\mcv_{\mathrm {L}}(E)$ will denote
in the following the empirical potential.

The use of two different energy shifts to treat the elastic and
inelastic parts of the hadronic tensor represents a complication. In
order to overcome this difficulty, it is sufficient to find a simple
way of calculating the elastic part.

We only deal with positive values of $E$, since we consider only
scattering states. Let $|\chi_{\mathrm {L,eff}}^{(-)}(E)\rangle$
denote the eigenvector of $h_{\mathrm {L,eff}}^{\dagger}(E)$, related
to the eigenvalue $E$, and satisfying the condition of incoming
scattering wave. The right hand side of Eq.~(\ref{eq:3.108}) can be
handled to obtain
\begin{equation}
\mcs_{\mathrm {L,eff}}^{\mathrm{el}}(E) = \sum |\chi_{\mathrm
{L,eff}}^{(-)}(E)\rangle \langle \chi_{\mathrm{L,eff}}^{(-)}(E)| ,
\label{eq:3.112}
\end{equation}
where the sum refers to the understood degeneracy indices. The
nontrivial proof of Eq.~(\ref{eq:3.112}) can be found in Sec.~4.12
of Ref.~\cite{capma2}. We remark that this proof extends, without any
modification, to a general Green's function and its self-energy.

Equation (\ref{eq:3.112}) involves the eigenvectors $|\chi_{\mathrm
{L,eff}}^{(-)}(E)\rangle$ of $h_{\mathrm {L,eff}}^{\dagger}(E)$,
which is a complicated Hamiltonian, but can be expressed in terms of
the corresponding eigenvectors $|\chi_{\mathrm {L}}^{(-)}(E)\rangle$
of $h_{\mathrm {L}}^{\dagger}(E)$. In fact, by a simple substitution
in the eigenvalue equation for $h_{\mathrm {L,eff}}^{\dagger}(E)$ and
using Eq.~(\ref{eq:3.111}), one checks that these eigenvectors are
related by the phase-conserving relation:
\begin{equation}
|\chi_{\mathrm {L,eff}}^{(-)}(E)\rangle = [ 1 - (\mcv'_{\mathrm
{L}})^{\dagger}(E)]^{1/2} |\chi_{\mathrm{L}}^{(-)}(E) \rangle.
\label{eq:3.113}
\end{equation}
Therefore, Eq.~(\ref{eq:3.112}) reads
\begin{equation}
\mcs_{\mathrm {L,eff}}^{\mathrm{el}}(E) = \sum [ 1 - (\mcv'_{\mathrm
{L}})^{\dagger}(E)]^{1/2} |\chi_{\mathrm {L}}^{(-)}(E)\rangle \langle
\chi_{\mathrm{L}}^{(-)}(E)| [ 1 - (\mcv'_{\mathrm {L}})(E)]^{1/2} .
\label{eq:3.114}
\end{equation}

From this equation, the elastic contribution to the hadronic tensor
can be calculated with the same difficulty as for the calculation of
an integrated single-proton knockout. In contrast, no similar way for
directly calculating the inelastic contribution is available. In
order to overcome this difficulty, the hadronic tensor can be
written as
\begin{equation}
W^{\mu\mu} (\omega, \q) = W_1^{\mu\mu} (\omega, \q) + W_2^{\mu\mu}
(\omega, \q) \label{eq:3.115}
\end{equation}
with
\begin{eqnarray}
& & W_1^{\mu\mu} (\omega , \q) = {\mathrm {Tr}} \, \big [ Kj^{\mu
\dagger} (\q) \mcs_{\mathrm {L,eff}} (E) j^\mu (\q) \big]
 \nonumber \\ & = & - \frac {1} {\pi} {\mathrm {Im}}
{\mathrm {Tr}} \big\{ Kj^{\mu \dagger} (\q)[ 1 - (\mcv'_{\mathrm
{L}})(E)]^{1/2} \mcg_{\mathrm {L}} (E)[ 1 - (\mcv'_{\mathrm
{L}})(E)]^{1/2}j^\mu (\q)\big\} , \label{eq:3.116}
\end{eqnarray}
and
\begin{equation}
W_2^{\mu\mu} (\omega , \q) = {\mathrm {Tr}} \, \big\{ Kj^{\mu
\dagger} (\q) [\mcs_{\mathrm {L,eff}}^{\mathrm {el}} (E -
\bar\epsilon) - \mcs_{\mathrm {L,eff}}^{\mathrm {el}} (E)] j^\mu (\q)
\big\} , \label{eq:3.117}
\end{equation}
where $E = \omega + E_0 -\epsilon_0$.

In the following, $W_2^{\mu\mu} (\omega , \q)$ will be calculated
using Eq.~(\ref{eq:3.114}), and $W_1^{\mu\mu} (\omega , \q)$ will be
obtained from the method of Ref.~\cite{capuzzi}, based on the
spectral representation of the s.p. Green's function.

\section{Spectral representation of the hadronic tensor}
\label{sec.spectral}

In this Section we consider the spectral representation of the s.p.
Green's function which allows practical calculations of the hadron
tensor of Eq.~(\ref{eq:3.116}). Due to the complex nature of
$\mcv_{\mathrm L}(E)$, the spectral representation of $\mcg_{\mathrm
L}(E)$ involves a biorthogonal expansion in terms of the
eigenfunctions of $h_{\mathrm L}(E)$ and $h_{\mathrm
L}^{\dagger}(E)$. We consider the incoming wave scattering solutions
of the eigenvalue equations, i.e.,
\begin{eqnarray}
& & \left(\mathcal{E} - h_{\mathrm L}^{\dagger}(E)\right) \mid
\chi_{\mathrm L,\mathcal{E}}^{(-)}(E)\rangle = 0 \ , \label{eq.inco1}
\\
& & \left(\mathcal{E} - h_{\mathrm L}(E)\right) \mid \tilde
{\chi}_{\mathrm L,\mathcal{E}}^{(-)}(E)\rangle = 0 \ .
\label{eq.inco2}
\end{eqnarray}
The choice of incoming wave solutions is not strictly necessary, but
it is convenient in order to have a closer comparison with the
treatment of the exclusive reactions, where the final states fulfill
this asymptotic condition and are the eigenfunctions
$\mid\chi_{\mathrm L,E}^{(-)}(E)\rangle$ of $h_{\mathrm
L}^{\dagger}(E)$.

The eigenfunctions of Eqs.~(\ref{eq.inco1}) and (\ref{eq.inco2})
satisfy the biorthogonality condition
\begin{equation}
\langle\chi_{\mathrm L,\mathcal{E}}^{(-)}(E)\mid \tilde
{\chi}_{\mathrm L,\mathcal{E}'}^{(-)}(E)\rangle = \delta
\left(\mathcal{E} - \mathcal{E}' \right) \ , \label{bicon}
\end{equation}
and, in absence of bound eigenstates, the completeness relation
\begin{equation}
\int_0^{\infty} \diff \mathcal{E}\mid\tilde {\chi}_{\mathrm
L,\mathcal{E}}^{(-)}(E)\rangle\langle \chi_{\mathrm
L,\mathcal{E}}^{(-)}(E)\mid =1 . \label{eq.comple}
\end{equation}

Equations (\ref{bicon}) and (\ref{eq.comple}) are the mathematical
basis for the biorthogonal expansions. The contribution of possible
bound state solutions of Eqs.~(\ref{eq.inco1}) and (\ref{eq.inco2})
can be disregarded in Eq.~(\ref{eq.comple}) since their effect on the
hadron tensor is negligible at the energy and momentum transfers
considered in this paper.

Using Eqs.~(\ref{eq.comple}) and (\ref{eq.inco2}), one obtains the
spectral representation
\begin{equation}
\mcg_{\mathrm L}(E) = \int_0^{\infty} \diff \mathcal{E}\mid\tilde
{\chi}_{\mathrm L,\mathcal{E}}^{(-)}(E)\rangle
\frac{1}{E-\mathcal{E}+i\eta} \langle\chi_{\mathrm
L,\mathcal{E}}^{(-)}(E)\mid \ . \label{eq.sperep}
\end{equation}
Therefore, $W_1^{\mu\mu} (\omega , \q)$ (Eq.~(\ref{eq:3.116})) can be
written, after expanding the ODM $K$ in terms of the natural
orbitals, as
\begin{eqnarray}
W_1^{\mu\mu}(\omega , \q) = -\frac{1}{\pi} \sum_{\nu} \mathrm{Im}
\bigg[  \int_0^{\infty}  & \diff \mathcal{E} & \frac{1}{\omega + E_0 -
\epsilon_0-\mathcal{E}+i\eta} \nonumber\\
& \times &  T_{\nu}^{\mu\mu} (\mathcal{E} ,\omega + E_0 -
\epsilon_0) \bigg]   \
, \label{eq.pracw}
\end{eqnarray}
where
\begin{eqnarray}
T_{\nu}^{\mu\mu}(\mathcal{E} ,E) & = & n_{\nu}\langle u_{\nu} \mid
j^{\mu\dagger}(\q) \sqrt{1-\mcv'_{\mathrm L}(E)}
\mid\tilde{\chi}_{\mathrm L,\mathcal{E}}^{(-)}(E)\rangle \nonumber \\
& \times &
 \langle\chi_{\mathrm L,\mathcal{E}}^{(-)}(E)\mid
 \sqrt{1-\mcv'_{\mathrm L}(E)} j^{\mu}(\q)\mid u_{\nu} \rangle \ .
\label{eq.tprac}
\end{eqnarray}
The quantities $|u_{\nu}\rangle$ and $n_\nu$ are the natural orbitals
and the occupation numbers, defined in Eq.~(\ref{eq:2.5}), and come
from the natural expansion of the ODM $K$ of Eq.~(\ref{eq:3.116}). The
limit for $\eta \rightarrow +0$, understood before the integral of
Eq.~(\ref{eq.pracw}), can be calculated exploiting the standard
symbolic relation
\begin{equation}
\lim_{\eta \rightarrow +0} \frac{1}{E-\mathcal{E}+i\eta} =
\mathcal{P} \left(\frac{1}{E-\mathcal{E}}\right) - i \pi \delta
\left(E-\mathcal{E}\right) \ , \label{eq.princ}
\end{equation}
where $\mathcal{P}$ denotes the principal value of the integral.
Therefore, Eq.~(\ref{eq.pracw}) reads
\begin{eqnarray}
W_1^{\mu\mu} (\omega , \q) & = & \sum_{\nu} \Bigg[ \mathrm{Re}
T_{\nu}^{\mu\mu} (\omega + E_0 - \epsilon_0, \omega + E_0 -
\epsilon_0)
\nonumber \\
& - & \frac{1}{\pi} \mathcal{P} \int_0^{\infty} \diff \mathcal{E}
\frac{1}{\omega + E_0 -\epsilon_0-\mathcal{E}} \mathrm{Im}
T_{\nu}^{\mu\mu} (\mathcal{E},\omega + E_0 - \epsilon_0) \Bigg] \ ,
\label{eq.finale}
\end{eqnarray}
which separately involves the real and imaginary parts of
$T_{\nu}^{\mu\mu}$.

The contribution coming from Eq.~(\ref{eq:3.117}) can be calculated
using Eq.~(\ref{eq:3.114}), i.e. as
\begin{eqnarray}
W_{2}^{\mu\mu}(\omega, \q) & = & \sum _{\nu}n_{\nu}\big [\langle
u_{\nu} | j^{\mu\dagger}(\q) \sqrt{1-\mcv'_{\mathrm L}(\bar E)}
|{\chi}_{\mathrm L,\bar E}^{(-)} (\bar E)\rangle \nonumber \\
&\times& \langle\chi_{\mathrm L,\bar E}^{(-)}(\bar E)|
\sqrt{1-\mcv'_{\mathrm L}(\bar E)} j^{\mu}(\q)| u_{\nu} \rangle
\nonumber \\
& - & \langle u_{\nu} | j^{\mu\dagger}(\q) \sqrt{1-\mcv'_{\mathrm
L}(E)} |{\chi}_{\mathrm L, E}^{(-)}(E)\rangle \nonumber \\
&\times& \langle\chi_{\mathrm L,E}^{(-)}(E)|
\sqrt{1-\mcv'_{\mathrm L}(E)} j^{\mu}(\q)| u_{\nu} \rangle \big ] \
, \label{eq.w2}
\end{eqnarray}
with $E = \omega + E_0 - \epsilon_0$ and $\bar E = E -\bar \epsilon$.
In this equation, only the optical potential wave functions of
Eq.~(\ref{eq.inco1}) appear everywhere, and the calculation is similar
to that of the integral of the exclusive cross section, but for the
different energy values which are involved and the presence of
occupation numbers and natural orbitals.

Some remarks on Eq.~(\ref{eq.finale}) are in order. Let us disregard
the square root correction, due to interference effects, and the
minor contribution of the integral over the energy. One has
\begin{equation}
W_1^{\mu\mu} (\omega , \q) \simeq \sum_{\nu} n_{\nu} \mathrm{Re}
\big [ \langle \chi_{\mathrm L,E}^{(-)}(E) | j^{\mu}(\q) |
u_{\nu}\rangle \langle u_{\nu}| j^{\mu\dagger}(\q) |{\tilde
\chi}_{\mathrm L,E}^{(-)}(E)\rangle \big ] . \label{eq:165}
\end{equation}
Let us compare Eq.~(\ref{eq:165}) with the corresponding elastic
contribution due to $\mcs_{\mathrm {L}}^{\mathrm{el}}(E)$, given by
Eq.~(\ref{eq:3.114}) without the square root corrections, i.e.
\begin{eqnarray}
W_1^{{\mathrm{el}} \, \mu\mu} (\omega , \q) = \sum_{\nu} n_{\nu}
\langle \chi_{\mathrm L,E}^{(-)}(E) | j^{\mu}(\q) | u_{\nu}\rangle
\langle u_{\nu}| j^{\mu\dagger}(\q) |{\chi}_{\mathrm
L,E}^{(-)}(E)\rangle . \label{eq:166}
\end{eqnarray}
In Eq.~(\ref{eq:166}) the attenuation of the strength, mathematically
due to the imaginary part of $\mcv_{\mathrm L}^{\dagger}(E)$, is
related to the flux lost toward the inelastic channels. In the
inclusive response this attenuation must be compensated by a
corresponding gain due to the inelastic contribution to $W_1^{\mu\mu}
(\omega , \q)$. In the description provided by Eq.~(\ref{eq:165}),
including both elastic and inelastic contributions, the attenuation
of strength of the first factor in the square bracket is compensated
by the second factor, where the imaginary part of $\mcv_{\mathrm
L}(E)$ has the effect of increasing the strength. We want to stress
here that in the Green's function approach it is just the imaginary
part of the optical potential which accounts for the redistribution
of the strength among different channels.

The main difference of the present approach with the one of
Ref.~\cite{capuzzi} is that now we are able to include in the initial
states the effect of correlations. In Ref.~\cite{capuzzi}, indeed,
the initial states were calculated as the overlaps between the target
nucleus and the residual nucleus described as a hole state,
corresponding to the separation energy taken from phenomenology, or
computed through an independent-particle model. Here, the initial
states are described through a realistic ODM, including correlations.
However, this goal is obtained using a constant separation energy,
i.e. the one corresponding to the ground state of the residual
nucleus, and this produces an undue shift of the cross section. The
contribution of Eq.~(\ref{eq:3.117}) is added in order to compensate
the shift.

\section{Realistic one-body density and correlations
\label{sec.density}}

The main issue of this paper is, together with a new and completely
antisymmetrized presentation of the Green's function approach, to
investigate the effect of nuclear correlations in the inclusive
quasielastic electron scattering. Correlations are included in the
ODM, which is expressed within the natural orbital representation
\cite{Low55} in the form
\begin{equation}
K(\mathbf{r},\mathbf{r}') \equiv \langle\r|K|\rf\rangle = 
 \sum_\nu n_\nu u^*_\nu(\r') u_\nu(\r).
\label{eq:density}
\end{equation}

The calculations presented in the next Section are done and compared 
for different realistic density
matrices which include the contribution of short-range and tensor
correlations and are obtained within different approaches. In
particular, we consider the Jastrow correlation method (JCM)
\cite{SAD93}, the correlated basis function (CBF) method
\cite{NDD+97}, and the Green's function method (GFM) \cite{PMD96}.

%\subsection{The Jastrow correlation method}

In Ref.~\cite{SAD93} the ODM is obtained within the JCM in its low-order
approximation (LOA).
The JCM incorporates the nucleon-nucleon (NN)
short-range correlations (SRC) starting from the ansatz for 
the wave-function of $A$ fermions \cite{Jas55}:
\begin{equation}
\Psi ^A(\mathbf{r}_1,\mathbf{r}_2,\ldots ,\mathbf{
r}_{A})=(C_{A})^{-1/2} \prod_{1\leq i<j\leq A} F(\mid \mathbf{
r}_{i}-\mathbf{r}_{j}\mid ) \Phi^{A}( \mathbf{r}_1,\mathbf{
r}_2,\ldots ,\mathbf{r}_{A}), \label{eq:jas1}
\end{equation}
where $C_A$ is a normalization constant and $\Phi^A$ is a single
Slater determinant wave function built from harmonic-oscillator (HO) s.p.
wave functions which depend on the oscillator parameter
$\alpha_{osc.}$, having the same value for both protons and neutrons.
Only central correlations are included in the correlation factor
$F(r)$, which is state-independent and has a simple Gaussian form
\begin{equation} \label{eq:jas2}
F(r) = 1- \exp (-\beta ^{2}r^{2}),
\end{equation}
where the correlation parameter $\beta $ determines the healing
distance. The LOA \cite{GGR71} keeps all terms up to the second order
in $(F-1)$ and the first order in $(F^2-1)$ in such a way that the
normalization of the density matrices is ensured order by order.
Under the above assumptions analytical expressions for the ODM and
corresponding natural orbitals have been obtained in \cite{SAD93}.
The values of the parameters $\alpha_{osc.}$ and $\beta $ have been
obtained \cite{SAD93} phenomenologically by fitting the experimental
elastic form factor data. Thus, in the present calculations the
following values of the parameters are used: $\alpha_{osc.}=0.59$
fm$^{-1}$, $\beta =1.43$ fm$^{-1}$ for $^{16}$O and
$\alpha_{osc.}=0.52$ fm$^{-1}$, $\beta =1.21$ fm$^{-1}$ for
$^{40}$Ca, respectively.

%\subsection{The CBF theory}

In the CBF method the trial $A$-particle wave function has the form
\begin{equation} \label{eq:vn1}
\Psi^A (x_{1},...,x_{A})=\mathcal{S}\left[ \prod_{i<j=1}^{A} F
(x_{i},x_{j})\right] \Phi^A (x_{1},...,x_{A}),
\end{equation}
where $x_i$ are particle coordinates which contain spatial, spin, and
isospin variables, $\mathcal{S}$ is a symmetrization operator, and
$\Phi^A$ is an uncorrelated (Slater determinant) wave function
normalized to unity and describing a closed-shell spherical system.
The correlation factor $F$ is generally written as
\begin{equation} \label{eq:vn2}
F(x_{i},x_{j})=\sum_{n}h_{n}(|\mathbf{r}_{i}-{\bf
r}_{j}|)\hat{\mathcal{O}}_n
\end{equation}
with basic two-nucleon operators $\hat{\mathcal{O}}_n$ inducing
central, spin-spin, tensor and spin-orbit correlations, either with
or without isospin exchange.
The ODM generated by a CBF-type correlated wave function for $^{16}$O
has been constructed in \cite{NDD+97}. The LOA,
which keeps terms up to the first order of the function
$h(x_{i},x_{j};x_{i}',x_{j}')= F(x_{i}',x_{j}') F(x_{i},x_{j})-1$, has
been used. The s.p. orbits entering the Slater determinant
$\Phi^A$ are taken from a Hartree-Fock calculation with the
Skyrme-III effective force. The correlation factor $F(x_{1},x_{2})$
obtained in \cite{PWP92} by variational Monte Carlo calculations with
Argonne $v_{14}$ NN forces has been used. The two-nucleon correlation
factors were restricted to the central, spin-isospin and
tensor-isospin operators.

%\subsection{The Green function approach}

For a nucleus like $^{16}$O with $J=0$ ground state angular momentum
the ODM can easily be separated into submatrices
of a given orbital angular momentum $l$ and total angular momentum
$j$. Within the GFM \cite{DM92,MPD95} the ODM in
momentum representation can be evaluated from the imaginary part of
the s.p. Green's function by integrating
\begin{equation} \label{eq:m1}
K_{lj}(k_1,k_2)=\int_{-\infty }^{\varepsilon_{F}}dE\frac{1}{\pi}
\text{Im} (\mathcal{G}_{lj}(k_1,k_2;E)),
\end{equation}

where the energy variable $E$ corresponds to the energy difference
between the ground state of the $A$ particle system and the energies
of the states in the $(A-1)$-particle system (negative $E$ with large
absolute value correspond to high excitation energies of the residual
system) and $\varepsilon _{F}$ is the Fermi energy. The
s.p. Green's function $\mathcal{G}_{lj}$ is obtained from
the solution of the Dyson equation
\begin{eqnarray}\label{eq:m2}
\mathcal{G}_{lj}(k_1,k_2;E)&= & \mathcal{G}_{lj}^{(0)}(k_1,k_2;E)+
\int dk_3 \int dk_4 \mathcal{G}_{lj}^{(0)} (k_1,k_3;E) \nonumber \\
& \times & \Delta\Sigma_{lj}(k_3,k_4;E) \mathcal{G}_{lj}(k_4,k_2;E),
\end{eqnarray}
where $\mathcal{G}^{(0)}$ refers to the Hartree-Fock propagator and
$\Delta\Sigma _{lj}$ represents contributions to the real and
imaginary part of the irreducible self-energy, which go beyond the
Hartree-Fock approximation of the nucleon self-energy used to derive
$\mathcal{G}^{(0)}$. The results for the ODM have been analyzed in
\cite{PMD96} in terms of the natural orbitals and the occupation
numbers in $^{16}$O. Within the natural orbital
representation they can be determined by diagonalizing the ODM of 
the correlated system.
The numerical results from \cite{PMD96} show that the ODM can be
described quite accurately in terms of four natural orbitals for each
partial wave $lj$.

\section{Results and discussion \label{sec.results}}

The model presented in this paper has been applied to evaluate the response
functions of the inclusive quasielastic electron scattering for the
nuclei $^{16}$O and $^{40}$Ca. For the nucleus $^{16}$O, no data are
available for the separated longitudinal and transverse response
functions. However, we can compare the results produced by several
realistic ODM, obtained within different correlation methods. For the
nucleus $^{40}$Ca, experimental data are available and we can be
compare our results with them.

Calculations are performed in a nonrelativistic approach including
the contributions of both $W_1$ (Eq.~(\ref{eq.finale})) and $W_2$
(Eq.~(\ref{eq.w2})), but disregarding the second term in $W_1$. The
evaluation of this term, which contains the principal value, requires
the integration over all the eigenfunctions of the continuum spectrum
of the optical potential and represents a quite complicate task.
Moreover, it would be useless in a nonrelativistic approach where the
contribution of this term is very small~\cite{capuzzi}. In $W_1$ and
$W_2$ the sum over all the natural orbitals $u_\nu$ is involved and
partial occupation numbers $n_\nu$ are included in the model. All
the results presented in the following are normalized to the number
of nucleons in the nucleus.
The s.p. energies $\epsilon_n$ and the spectroscopic factors
$\lambda_n$ we have used in the calculation of the 
$W_2$ contribution are those corresponding to the overlap states obtained 
in Ref.~\cite{SAD96} for JCM, in \cite{NDD+97} for CBF, and in \cite{GPD+97}
for GFM, following the method of Ref.~\cite{NWH93}.

The other ingredients in the matrix elements of Eqs.~(\ref{eq.tprac})
and (\ref{eq.w2}) are the same as those taken in our previous
application of the Green's function approach to the inclusive
electron scattering (Ref.~\cite{capuzzi}) and in our treatment of the
exclusive one-nucleon knockout, which is based on the distorted wave
impulse approximation and was widely and successfully applied to the
analysis of $(e,e'p)$ data~\cite{book,DWEEPY}. The one-body nuclear
current operator is given by the nonrelativistic approximation of
Ref.~\cite{McVoy}, $\tilde \chi$ and ${\chi}$ are eigenfunctions of a
phenomenological spin-dependent optical potential and of its
hermitian conjugate, determined through a fit to elastic
nucleon-nucleus scattering data including cross sections and
polarizations~\cite{Schwandt}. This allows a consistent treatment of
FSI in the inclusive and in the exclusive electron scattering. The
role of FSI in the Green's function approach was already investigated
in Ref.~\cite{capuzzi} in a nonrelativistic framework and, more
recently, in Ref.~\cite{meucci} in a relativistic framework and will
not be discussed later on in this paper.

%%%%%%%%%%%%%%% Fig. 1%%%%%%%%%%%%%%%%%%%
\begin{figure}[ht]
\begin{center}
\includegraphics[height=15cm,width=12cm]{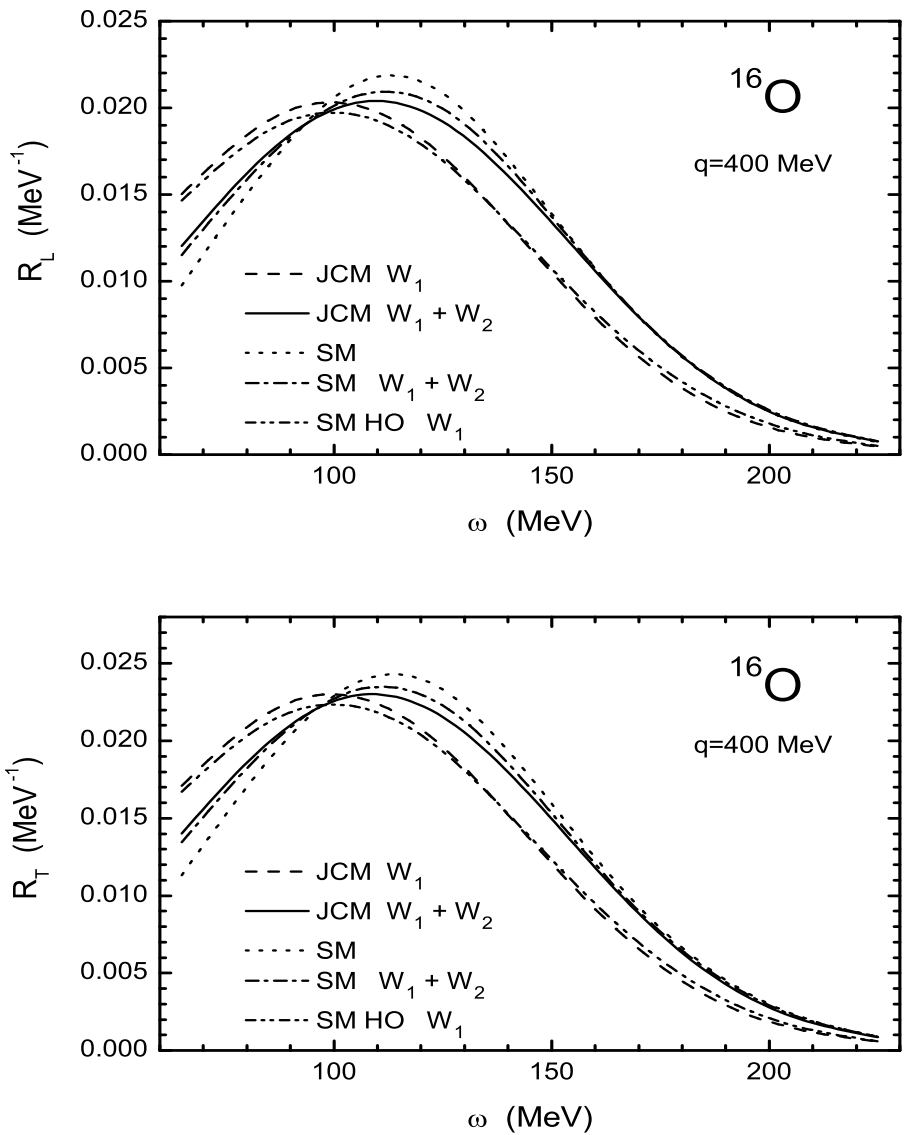}
\vskip -0.1cm
\caption {Longitudinal (upper panel) and transverse
(lower panel) response functions for the $^{16}$O$(e,e')$ reaction at
$q = 400$ MeV$/c$. The curves indicated by JCM $W_1$ and JCM
$W_1+W_2$ give the contributions of $W_1$ and of the sum $W_1+ W_2$
(see Eqs.~(\ref{eq.finale}) and (\ref{eq.w2})) with the ODM of the
JCM~\cite{SAD93}. SM is obtained using the SM prescription of
Ref.~\cite{capuzzi} with the phenomenological s.p. wave functions of
Ref.~\cite{ES}. SM $W_1+W_2$ is the sum of the two terms $W_1$ and
$W_2$ where the ODM is replaced by the SM prescription
(Eq.~(\ref{eq.SM2})) with the phenomenological s.p. wave functions. SM
HO $W_1$ gives the contribution of $W_1$ in the SM with harmonic
oscillator wave functions.} \label{fig1}
\end{center}
\end{figure}
%%%%%%%%%%%%%%%%%%%%%%%%%%%%%%%%%%%%%%%%%%%%%%%%%%%%
A numerical example for the longitudinal and transverse response
functions of the $^{16}$O$(e,e')$ reaction at $q = 400$ MeV$/c$ is
given in Fig.~\ref{fig1}. The curves called JCM $W_1$ and JCM
$W_1+W_2$ give the contribution of $W_1$ and of the sum $W_1+W_2$ and
are calculated with the ODM within the JCM in its LOA, 
where only short-range correlations are included. The
SM curve corresponds to the prescription adopted in our previous
applications of the Green's function approach~\cite{capuzzi}, where
correlations are neglected and the components of the hadronic tensor
are given by
\begin{equation}
W^{\mu\mu} (\omega , \q) = \sum_{n} \mathrm{Re} T_{n}^{\mu\mu}
(\omega + E_0 - \epsilon_n, \omega + E_0 - \epsilon_n), \label{eq.SM}
\end{equation}
with
\begin{eqnarray}
T_{n}^{\mu\mu}(E,E) & = & \lambda_{n}\langle \varphi_{n} \mid
j^{\mu\dagger}(\q) \sqrt{1-\mcv'_{\mathrm L}(E)}
\mid\tilde{\chi}_{\mathrm L,E}^{(-)}(E)\rangle \nonumber \\
& \times &
 \langle\chi_{\mathrm L,E}^{(-)}(E)\mid
 \sqrt{1-\mcv'_{\mathrm L}(E)} j^{\mu}(\q)\mid \varphi_{n} \rangle \
. \label{eq.SM1}
\end{eqnarray}
A pure shell model (SM) is assumed for the nuclear structure: the sum
is over all the occupied states in the SM and a unitary spectral
strength ($\lambda_n=1$) is taken for each s.p. state $\varphi_n$. In
the calculations for the SM curve the wave functions $\varphi_n$ are
taken from a phenomenological Woods-Saxon potential ~\cite{ES} and
$\epsilon_n$ are the experimental excitation energies of the states
$n$.

The difference between the correlated JCM $W_1+W_2$ and the
uncorrelated SM results indicates that correlations give a
redistribution of the strength (the total strength is conserved in
all the calculations presented in this paper) and a reduction of the
response functions by $\sim 10\%$ in the peak region. Part of the
difference is however due to the different methods. If the sum
$W_1+W_2$ is calculated in a SM approach, where the ODM is replaced
by
\begin{equation}
\langle\r|K|\rf\rangle = \sum_{n} \lambda_n \varphi^*_n(\r')
\varphi_n(\r), \label{eq.SM2}
\end{equation}
the corresponding result SM $W_1+W_2$, which is also displayed in
the figure, lies between the SM and JCM $W_1+W_2$ curves, and the
reduction produced by SRC in the peak region is
no more than $\sim 5\%$. 
The different position of the peaks of the curves JCM $W_1+W_2$ 
and SM is mainly due to the following reasons. i) In the JCM $W_1+W_2$ 
curve the inelastic part of the spectral function $\mcs_n$ is calculated using 
the total energy prescription of Eq.~(\ref{eq:3.82}) which does not produce a
shift, differently from the kinetic energy prescription (\ref{eq:3.81}) 
adopted for the SM curve. The effect of these different prescriptions is
indicated by the distance between the peaks of SM $W_1+W_2$ and SM. 
ii) In the cases JCM $W_1+W_2$ and SM the calculations
use different values of $\epsilon_n$. The resulting effect is indicated 
by the distance between the peaks of JCM $W_1+W_2$ and SM $W_1+W_2$.

The term $W_2$ gives a shift of the calculated response functions
that is important to determine the position of the peak and to bring
it close to the SM result. In the previous application of
Ref.~\cite{capuzzi} the SM result was able to give a fair description
of the size and shape of the experimental longitudinal response of
the $^{12}$C$(e,e')$ reaction, in a range of momentum transfer
between 400 and 550 MeV$/c$. In contrast, the experimental transverse
response was generally underestimated. For the transverse response,
however, an important contribution might be given by two-body
meson-exchange currents, which are not included in the present model
based on the s.p. Green's function.

The last curve drawn in Fig.~\ref{fig1} gives the contribution of
$W_1$ calculated in the SM approach (Eq.~(\ref{eq.SM2})) and with the
HO single-particle wave functions used in the
calculation of the ODM with the JCM (SM HO $W_1$). The comparison
with the JCM $W_1$ result indicates that correlation effects are
still within $\sim 5\%$, but in this case they enhance the response
functions in the peak region. Phenomenological Woods-Saxon wave
functions certainly represent a more reliable basis for a SM
calculation. A calculation with HO wave functions, however, may allow
a more consistent comparison with the correlated ODM of the JCM and
may give an idea of the uncertainty produced by different s.p. wave
functions.

The comparison between the uncorrelated SM and the correlated JCM
results shows that the effects of correlations depend on the
uncorrelated result that is considered for the comparison. The
effects of SRC are however small and within the
range of uncertainty produced by the choice of the theoretical
ingredients. From this point of view our results are in substantial
agreement with those of Ref.~\cite{Co}, where the effects of
SRC on the response functions of the
quasielastic electron scattering were investigated in a different
model. In Fig.~\ref{fig1} correlation effects have a similar behavior
on the longitudinal and transverse responses, while in Ref.~\cite{Co}
correlations act differently on the two responses. This apparent
discrepancy might be explained by the different models used here and
in Ref.~\cite{Co}, by the small effects produced by correlations, and
by the uncertainties related to the other ingredients of the
calculations.

%%%%%%%%%%%%%%% Fig. 2%%%%%%%%%%%%%%%%%%%
\begin{figure}[ht]
\begin{center}
\includegraphics[height=15cm,width=12cm]{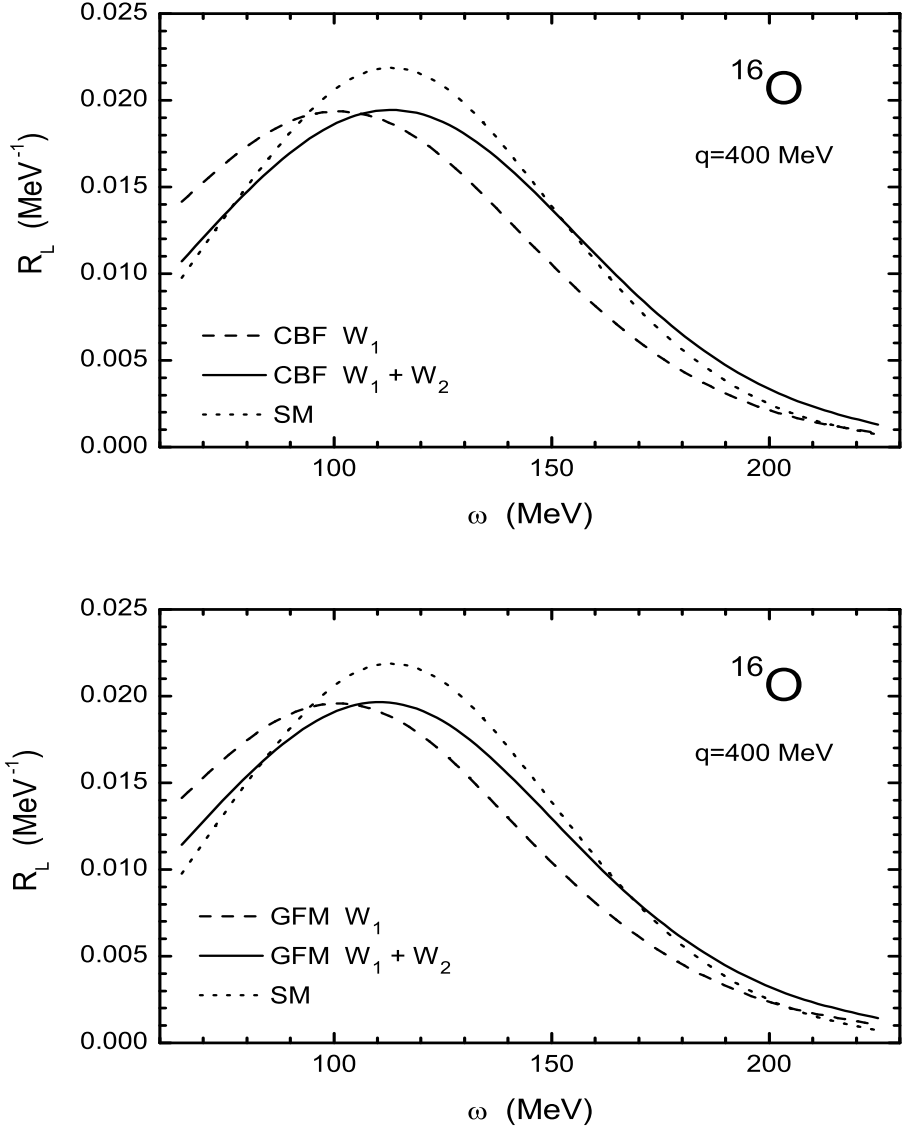}
\vskip -0.1cm
\caption {Longitudinal response functions for the
$^{16}$O$(e,e')$ reaction at $q = 400$ MeV$/c$. The contributions of
$W_1$ and of the sum $W_1+ W_2$ with the ODM of the
CBF~\cite{NDD+97} and of the GFM~\cite{PMD96} are displayed in the
upper and lower panels, respectively. The SM result is the same as in
Fig.~\ref{fig1}.} \label{fig2}
\end{center}
\end{figure}
%%%%%%%%%%%%%%%%%%%%%%%%%%%%%%%%%%%%%%%%%%%%%%%%%%%%
The longitudinal response functions of the $^{16}$O$(e,e')$ reaction
at $q = 400$ MeV$/c$ obtained with the ODM within the
CBF~\cite{NDD+97} and the GFM~\cite{PMD96} are shown in Fig.
\ref{fig2}. In both calculations the ODM includes both short-range
and tensor correlations. The two density matrices, which are obtained
within different correlation methods, give similar results. In
comparison with the SM curve, correlations produce a redistribution
of the strength and a reduction of the response in the peak region by
$\sim 15\%$. In comparison with the SM $W_1+ W_2$ result (that is
not shown Fig.~\ref{fig2}, but is given in Fig.~\ref{fig1}) the
reduction is within $10\%$, but anyhow larger than for the JCM
density matrix, where only SRC are included. The
term $W_2$ gives a shift of the response function that brings the
position of the maximum toward the SM result.

%%%%%%%%%%%%%%% Fig. 3%%%%%%%%%%%%%%%%%%%
\begin{figure}[ht]
\begin{center}
\includegraphics[height=8cm,width=12cm]{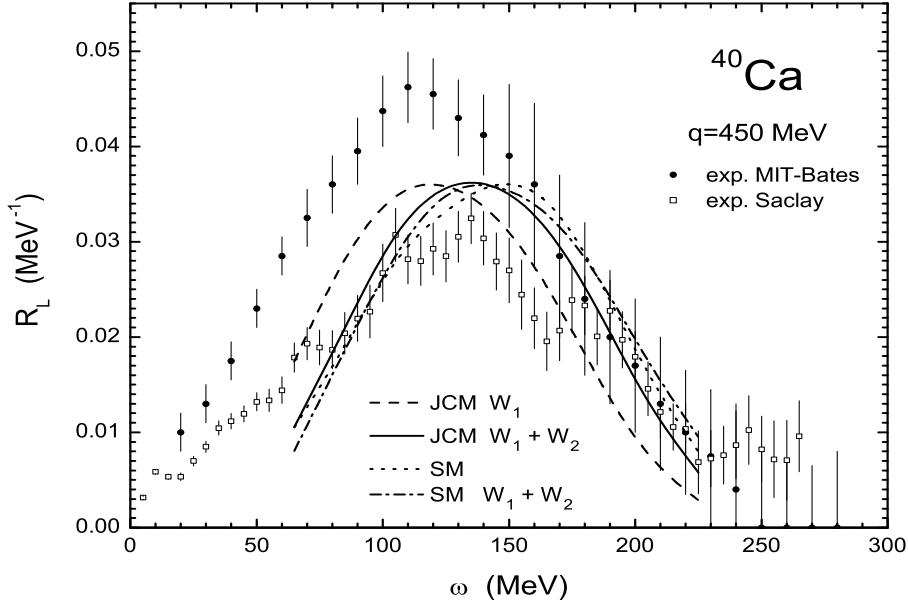}
\vskip -0.1cm
\caption {Longitudinal response functions for the
$^{40}$Ca$(e,e')$ reaction at $q = 450$ MeV$/c$. The contributions of
$W_1$ and of the sum $W_1+W_2$ with the ODM of the
JCM~\cite{SAD93} are displayed together with the SM and SM $W_1+
W_2$ results produced by the phenomenological s.p. wave functions of
Ref.~\cite{ES}. The Saclay data (open squares) are from
Ref.~\cite{meziani}, the MIT-Bates (black circles) are from
Ref.~\cite{batesca}. } \label{fig3}
\end{center}
\end{figure}
%%%%%%%%%%%%%%%%%%%%%%%%%%%%%%%%%%%%%%%%%%%%%%%%%%%%
The longitudinal response functions of the $^{40}$Ca$(e,e')$ reaction
at $q = 450$ MeV$/c$ is shown in Fig.~\ref{fig3}. The JCM $W_1$ and
JCM $W_1+W_2$ results are compared with the SM and SM $W_1+W_2$ ones
and with the available data~\cite{meziani,batesca}. The main
difference between the different results is in the position of the
maximum. The contribution of $W_2$ shifts the response by $\sim$ 15
MeV at larger values of the energy transfer. The difference between 
the SM $W_1+W_2$ and SM results and between the JCM $W_1+W_2$
and SM $W_1+W_2$ results indicates the sensitivity of the position of the
peak to the different prescriptions used for the energy shifts and to
the values of $\epsilon_n$ used in the calculations. These effects
are more important in a heavier nucleus like $^{40}$Ca than in
$^{16}$O. In comparison with data, our results are closer to the
Saclay data.

%%%%%%%%%%%%%%%%%%%%%%%%%%%%%%%%%%%%%%%%%%%%%%%
\section{Summary and conclusions \label{conc}}

In this paper we have presented a completely antisymmetrized Green's
function approach to inclusive quasielastic electron scattering.
The main goals are the following.

i) To express the final state interaction in terms of the
self-energy, rather than in terms of the Feshbach optical potential
as was done in previous papers, since the mass operator is more
closely related to the empirical optical-model potentials.
Therefore, we have developed a theoretical approach based on extended 
projection operators, as in Ref.~\cite{capma}.

ii) To include an approximated treatment of the interference between
different reaction channels, which is usually disregarded. 
This has required some modifications of the treatment of Ref.~\cite{capma},
resulting in a different separation of the hadronic tensor into a 
direct and an interference part. The latter one is essential to explain the
replacement of the self-energy with the empirical potential.

iii) To separate the elastic and inelastic contributions to the
hadronic tensor, in order to introduce different approximations
concerning the dependence on the state of the residual nucleus.

iv) To include in the calculation the effects of correlations in the
target nucleus through the use of a realistic one-body density
matrix.

The method has been applied to the reactions $^{16}$O($e,e'$) and
$^{40}$Ca($e,e'$). Results produced by density matrices including
short-range as well as tensor correlations have been compared. The
effects of short range correlations are small (within $\sim 5\%$) and
within the range of uncertainty related to the choice of other
ingredients of the calculation (e.g. the s.p. wave functions). This
result is in substantial agreement with previous
investigations~\cite{Co}, but a similar effect is found in the
present work on the longitudinal and the transverse response
functions. Stronger effects are obtained when also tensor
correlations are included. For the $^{16}$O($e,e'$) reaction at $q =
450$ MeV$/c$ the height of the peak is reduced by $\sim 10\%$ when
both types of correlations are considered. A correction dependent on
the s.p. energies is necessary [the term $W_2$ in Eq.~(\ref{eq.w2})]
in order to determine the position of the peak. 
The comparison of the present results
with the available experimental data for the nucleus $^{40}$Ca gives
a better agreement, in the longitudinal response, with the Saclay
data than with the MIT-Bates ones.

In order to get deeper insight into the dependence of the inclusive
electron scattering on correlations, one should also compute the
contribution of two-body currents and their interference with tensor
correlations, that seems to give a larger
effect\cite{Leidemann,Fabrocini,Sick}. In order to accomplish this
task, however, a new approach, based on the two-particle Green's
function and including the two-body density matrix, should be
developed.

\appendix
\section{APPENDIX}

A real number $a$ belongs to the continuous spectrum of a
self-adjoint operator $A$ if and only if:

i) it does not belong to the discrete spectrum,

ii) for every $\eta>0$, one can find a normalizable vector
$|\psi_{a,\eta}\rangle$ and a constant $C_a$ such that
\begin{equation}
\| (a - A)|\psi_{a,\eta}\rangle \|^2 \leq C_a \||\psi_{a,\eta}\rangle
\|^2 \eta . \label{eq:a.1}
\end{equation}

The vectors $|\psi_{a,\eta}\rangle$ are called {\lq\lq approximate
eigenvectors of $A$ related to the eigenvalue $a$\rq\rq} [see
Sec.~8.1 of Ref~\cite{richt}]. Now, we construct a set of
approximate eigenvectors of the Hamiltonian $H$ of the residual
nucleus, which, added to the exact bound eigenvectors $|n\rangle$,
satisfy a completeness relation in the limit $\eta \rightarrow +0$.
We remark that this property is not an automatic consequence of
Eq.~(\ref{eq:a.1}).

Let $\mch_c$ be the Hilbert subspace spanned by the non-normalizable
eigenvectors $|\epsilon\rangle$ of the continuous spectrum of $H$
which, for simplicity, is assumed to coincide with $[0,+\infty)$. Let
$\{|u_k\rangle\}$ be a basis in $\mch_c$ consisting of normalizable
vectors such that $\langle\epsilon|u_k\rangle$ are functions of
$\epsilon$ bounded, continuous, and everywhere different from zero.
This can be realized weighing the eigenvectors $|\epsilon\rangle$ by
means of a complete orthonormal set of functions having the same
properties. We set
\begin{equation}
|\epsilon,k;\eta\rangle \equiv \sqrt{\frac {\eta} {\pi}} (\epsilon -
H + i\eta)^{-1} |u_k\rangle . \label{eq:a.2}
\end{equation}
Note that the vectors $|\epsilon,k;\eta\rangle$ and $|n\rangle$ are
orthogonal.

\underline{Theorem 1}. The vectors $|\epsilon,k;\eta\rangle$ are
approximate eigenvectors of $H$.

\underline{Proof}. One has
\begin{equation}
\lim_{\eta \rightarrow +0} \| |\epsilon,k;\eta\rangle \|^2 =
|\langle\epsilon|u_k\rangle |^2 \neq 0 \label{eq:a.3}
\end{equation}
since
\begin{equation}
 \| |\epsilon,k;\eta\rangle \|^2 = \int_0^{\infty} \diff \, \epsilon'
 L(\epsilon - \epsilon'; \eta) |\langle\epsilon'|u_k\rangle |^2 \, ,
\,
 L(x; \eta) \equiv \frac {\eta} {\pi} \frac {1} {x^2 + \eta^2} ,
\label{eq:a.4}
\end{equation}
and, in the distributional sense of the limit,
\begin{equation}
\lim_{\eta \rightarrow +0} L(x; \eta) = \delta(x) . \label{eq:a.5}
\end{equation}
Moreover, the following inequality holds:
\begin{equation}
 \|(\epsilon - H) |\epsilon,k;\eta\rangle \|^2 =
 \frac {\eta} {\pi}\langle u_k| \frac {(\epsilon - H)^2} {(\epsilon -
H)^2 +
 \eta^2} |u_k \rangle \leq \frac {\eta} {\pi}\langle u_k|u_k \rangle
=
 \frac {\eta} {\pi} .
\label{eq:a.6}
\end{equation}
Therefore, one has
\begin{equation}
\lim_{\eta \rightarrow +0} \frac {\|(\epsilon - H)
|\epsilon,k;\eta\rangle \|^2} {\||\epsilon,k;\eta\rangle \|^2} = 0
\label{eq:a.7}
\end{equation}
and Eq.~(\ref{eq:a.1}) is satisfied. $\square$

\vskip 1cm

\underline{Theorem 2}. In the limit for $\eta \rightarrow +0$, one
has the completeness relation
\begin{equation}
\sum_n |n\rangle \langle n| + \lim_{\eta \rightarrow +0} \sum_k
\int_0^{\infty} \diff \epsilon \, |\epsilon,k;\eta\rangle \langle
\epsilon,k;\eta| = 1 , \label{eq:a.8}
\end{equation}
where the convergence is understood in the weak sense, i.e., inside a
scalar product between normalizable states.

\underline{Proof}. Due to the orthogonality between the vectors
$|\epsilon,k;\eta\rangle$ and $|n\rangle$, it is sufficient to prove
Eq.~(\ref{eq:a.8}) inside the scalar product between vectors
$|\phi\rangle$ and $|\chi\rangle$ belonging to $\mch_c$, where both
states $\{|\epsilon\rangle\}$ and $\{|u_k\rangle\}$ are complete.
Thus the sum over $n$ does not contribute, and one considers only
\begin{eqnarray}
& & \sum_k \int_0^{\infty}\diff \epsilon \, \langle
\phi|\epsilon,k;\eta\rangle \langle \epsilon,k;\eta|\chi\rangle =
\frac {\eta} {\pi} \int_0^{\infty} \diff \epsilon \, \langle \phi|
\frac {1} {(\epsilon - H)^2 +\eta^2}
|\chi \rangle \nonumber \\
& & = \int_0^{\infty}\diff \epsilon \, \int_0^{\infty}\diff
\epsilon' \, L(\epsilon - \epsilon';\eta) \langle
\phi|\epsilon'\rangle
\langle\epsilon'|\chi\rangle \nonumber \\
& & =\int_0^{\infty}\diff \epsilon' \, \langle \phi|\epsilon'\rangle
\langle\epsilon'|\chi\rangle \int_{-\epsilon'}^{\infty} \diff x \,
L(x;\eta) , \label{eq:a.9}
\end{eqnarray}
where in the last step we have exchanged the integrals, according to
the Fubini theorem, since $|L(\epsilon - \epsilon';\eta) \langle
\phi|\epsilon'\rangle \langle\epsilon'| \chi\rangle|$ is Lebesgue
summable in $\epsilon$ and $\epsilon'$. Observing that
\begin{eqnarray}
|\langle \phi|\epsilon'\rangle \langle\epsilon'|\chi\rangle
\int_{-\epsilon'}^{\infty} \diff x \, L(x;\eta) | & < & \, |\langle
\phi|\epsilon'\rangle \langle\epsilon'|\chi\rangle
\int_{-\infty}^{\infty} \diff x \, L(x;\eta) | \nonumber \\ & = &
\langle \phi|\epsilon'\rangle \langle\epsilon'|\chi\rangle ,
\label{eq:a.10}
\end{eqnarray}
where the last term is Lebesgue summable and does not depend on
$\eta$, we can use the dominated convergence theorem to take the
limit for $\eta\rightarrow +0$ within the first integral of the last
term in Eq.~(\ref{eq:a.9}). Thus, using Eq.~(\ref{eq:a.5}), one
obtains
\begin{eqnarray}
& & \lim_{\eta\rightarrow +0}\sum_k \int_0^{\infty}\diff \epsilon \,
\langle \phi|\epsilon,k;\eta\rangle\langle
\epsilon,k;\eta|\chi\rangle = \int_0^{\infty}\diff \epsilon'
\,\langle\phi|\epsilon'\rangle \langle\epsilon'|\chi\rangle
\int_{-\epsilon'}^{\infty} \diff x \,
\delta(x) \nonumber \\
& & =\int_0^{\infty}\diff \epsilon' \, \langle\phi|\epsilon'\rangle
\langle\epsilon'|\chi\rangle = \langle\phi|\chi\rangle \, , \,
\forall \, |\phi\rangle, |\chi\rangle \in \mch_c .\quad \square
\label{eq:a.11}
\end{eqnarray}

\vskip 1cm

Substituting the completeness relation of Eq.~(\ref{eq:3.14}) by
(\ref{eq:a.8}) and introducing the projection operators
\begin{equation}
P(\epsilon,k;\eta) \equiv \int \diff \p \, \alpha_{\mathbf p}
|\epsilon,k;\eta\rangle^{\mathrm N} \, ^{\mathrm N}\langle
\epsilon,k;\eta|\alpha_{\mathbf p} \, , \, |\epsilon,k;\eta
\rangle^{\mathrm N} \equiv \frac {|\epsilon,k;\eta\rangle}
{\||\epsilon,k;\eta\rangle \|}, \label{eq:a.12}
\end{equation}
the contribution of the continuous spectrum is recovered adding to
Eq.~(\ref{eq:3.22a}) the term
\begin{eqnarray}
& & \lim_{\eta\rightarrow +0}\sum_k \int\diff \epsilon \, \mathrm
{Re} \int \diff \p \diff \pf \, \langle \psi_0|a_{\mathbf
{p-q}}^{\dagger}|
\epsilon,k;\eta\rangle \nonumber \\
& \times & ^{\mathrm N}\langle\epsilon,k;\eta|\alpha_{\mathbf p}
\delta (\omega + E_0 -\epsilon - \hat H_{\epsilon})\alpha_{\mathbf
p'} | \epsilon,k;\eta\rangle^{\mathrm N} \, \,
\langle\epsilon,k;\eta|a_{\mathbf p'} J^0(\q)|\psi_0\rangle ,
\label{eq:a.13}
\end{eqnarray}
where $\hat H_{\epsilon}$ is defined analogously to $\hat H_n$ (see
Eq.~(\ref{eq:2.24})), i.e. as
\begin{equation}
\hat H_{\epsilon} = H - \epsilon \quad \mathrm{in} \quad \mch^{(Z+1)}
\quad , \quad \hat H_{\epsilon} = \epsilon - H \quad \mathrm{in}
\quad \mch^{(Z-1)} . \label{eq:a.14}
\end{equation}
Thus, using the approximation
\begin{equation}
^{\mathrm N}\langle\epsilon,k;\eta|\alpha_{\mathbf p} \delta (\omega
+ E_0 -\epsilon - \hat H_{\epsilon})\alpha_{\mathbf p'}
|\epsilon,k;\eta\rangle^{\mathrm N} \simeq \langle \p| \mcs_0 (\omega
+ E_0 -\epsilon_0 - \tilde\epsilon) |\pf\rangle, \label{eq:a.15}
\end{equation}
analogous to Eq.~(\ref{eq:3.34}), and using again Eq.~(\ref{eq:a.8})
one recovers Eqs.~(\ref{eq:3.37})--(\ref{eq:3.42}).

\begin{ack}

We would like to thank A. N. Antonov for useful
discussions and continuous encouragement. We are grateful to 
S. S. Dimitrova, H. M\"uther and D. Van Neck for the results 
of the ODM's we have used.

\end{ack}

%%%%%%%%%%%%%%%%%%%%%%%%%%%%%%%%%%%%%%%%%%%%%%%

\end{document}